\documentclass[aps,prc,twocolumn,superscriptaddress,showpacs]{revtex4}

\usepackage[dvips]{graphicx}
\usepackage{bm}
\usepackage{amsmath,amssymb}
\usepackage{hyperref}
\allowdisplaybreaks
\begin{document}

\title{Gamow-Teller strength distributions at finite temperatures and electron
capture in stellar environments}

\author{Alan~A.~Dzhioev}
\email[]{dzhioev@theor.jinr.ru}
\affiliation{Bogoliubov Laboratory of
Theoretical Physics, JINR, 141980, Dubna, Russia}
\author{A.~I.~Vdovin}
\affiliation{Bogoliubov Laboratory of Theoretical Physics, JINR,
141980, Dubna, Russia}
\author{V.~Yu.~Ponomarev}
\affiliation{Institut f{\"u}r Kernphysik, Technische Universit{\"a}t
  Darmstadt, 64289 Darmstadt, Germany}
\author{J.~Wambach}
\affiliation{Institut f{\"u}r Kernphysik, Technische Universit{\"a}t
  Darmstadt, 64289 Darmstadt, Germany}
\affiliation{GSI Helmholtzzentrum f\"ur Schwerionenforschung,
  Planckstr. 1, 64291 Darmstadt, Germany}
\author{K.~Langanke}
\affiliation{GSI Helmholtzzentrum f\"ur Schwerionenforschung,
  Planckstr. 1, 64291 Darmstadt, Germany}
\affiliation{Institut f{\"u}r Kernphysik, Technische Universit{\"a}t
  Darmstadt, 64289 Darmstadt, Germany}
\affiliation{Frankfurt Institute for Advanced Studies,
  Ruth-Moufang-Str. 1, 60438 Frankfurt, Germany}
\author{G.~Mart\'inez-Pinedo}
\affiliation{GSI Helmholtzzentrum f\"ur Schwerionenforschung,
  Planckstr. 1, 64291 Darmstadt, Germany}

\date{\today}

\begin{abstract}
We propose a new method to calculate stellar weak-interaction rates.
It  is  based on the Thermo-Field-Dynamics formalism and allows the
calculation of the weak-interaction response of nuclei at finite
temperatures. The thermal evolution of the GT$_+$ distributions is
presented for the sample nuclei $^{54,\,56}$Fe and~$^{76,78,80}$Ge.
For Ge we also calculate the strength distribution of
first-forbidden transitions. We show that thermal effects shift the
GT$_+$ centroid to lower excitation energies and make possible
negative- and low-energy transitions. In our model we demonstrate
that the unblocking effect for GT$_+$ transitions in neutron-rich
nuclei is sensitive to increasing temperature. The results are used
to calculate electron capture rates and are compared to those
obtained from the shell model.
\end{abstract}


\maketitle

\section{\label{introduction}Introduction}

The properties of nuclei at finite temperatures have attracted
attention for a long time. Due to a considerable amount of
experimental data the main subject of the study was the thermal
properties of the giant dipole resonance (see,
e.g.,~\cite{toro2000,shlomo2005} and reference therein).  In the
astrophysical context the thermal properties of Gamow-Teller (GT)
transitions are of special interest since they play a crucial role
in weak-interaction mediated reactions (electron capture,
beta-decay, neutrino scattering etc.)~\cite{lang03}. For example,
electron capture on iron group nuclei initiates the gravitational
collapse of the core of a massive star triggering supernova
explosions. Moreover, electron capture rates largely determine the
mass of the core and thus the fate of the shock wave formed by the
supernova explosion. Since the strong phase space dependence makes
the relevant stellar weak-interaction rates at the early stage of
the collapse -- when the electron chemical potential $\mu_e$ is of
the same order as the nuclear Q-values -- very sensitive to the GT
distributions these need to be calculated very accurately in this
regime. With proceeding collapse and hence increasing density,
$\mu_e$ grows faster than the Q-values of the nuclei present in the
matter composition, the capture rates become less sensitive to the
details of the GT distribution and are mainly determined by the
total GT strength and its centroid energy. However, forbidden
transitions can no longer be neglected when $\mu_e$ reaches values
of the order of 30 MeV at core densities $\rho > 10^{11}$
g/cm$^3$~\cite{coo84,lang03} The situation is further complicated by
the fact that weak-interaction processes in stellar environments
take place at temperatures of the order a few hundred keV to a few
MeV and GT transitions occur not only from the nuclear ground state,
but also from excited states.

From a microscopic point of view, there are two routes of handling
GT strength distributions and weak-interaction rates at finite
temperatures.  One is a state-by-state evaluation of the rate by
summing over Boltzmann-weighted, individually determined GT
strengths for the various states.  The second is based on an
equilibrium statistical formulation of the nuclear many-body
problem. In this approach, the thermal response of a nucleus to an
external perturbing field is given by the canonical (or grand
canonical) expectation value of the corresponding transition
operator. When applied to charge-exchange processes this method
yields the temperature-dependent GT and first forbidden strength
function that can be then used to calculate weak-interaction rates.

For $sd$ and $pf$-shell nuclei the first approach was originally
used by Fuller~et~al.~\cite{fuller} who calculated stellar
weak-interaction rates using the independent-particle shell-model,
supplemented by experimental data, whenever available. To allow for
GT transitions from nuclear excited states these authors employed
the Brink hypothesis. This assumes that the GT strength distribution
on excited states is the same as for the ground state, only shifted
by the excitation energy of the state. These rates were subsequently
updated, taking into account the quenching of axial coupling
constant~\cite{auf94}. Modern high-performance computing
capabilities combined with state-of-the-art diagonalization
approaches make possible shell-model calculations of the GT strength
distribution not only for the nuclear ground state, but also for the
few lowest excited states. It was demonstrated~\cite{lan98,lan99}
that even for the lowest excited states in the parent nucleus the
Brink hypothesis is valid only for the bulk of the GT strength, but
is not applicable for the individual transitions to states at
low-excitation energy in the daughter nucleus. Taking this into
account, the weak-interaction rates based on the shell-model
diagonalization approach~\cite{lan00,lan01_1} were derived from the
individual GT distributions from the lowest excited states and from
'back-resonant contributions' , i.e. from transitions determined
from the inverse GT distributions connecting excited states in the
daughter to the lowest states in the parent spectrum. However, the
compilation of~\cite{lan00,lan01_1} applied Brink's hypothesis when
taking into account GT transitions from highly excited states.
Weak-interaction rates have also been computed using the spectral
distribution theory~\cite{kar94} and the proton-neutron
quasiparticle RPA model~\cite{nab04}. The first method is also based
on the Brink hypothesis. The latter does not use this hypothesis,
but some uncertainties arise due to the approximate treatment of the
parent excited states as multi-quasiparticle states as well as the
insufficient knowledge of the quantum numbers of the states
involved. Recently stellar electron capture rates have been
calculated within the framework of the finite temperature RPA using
a set of Skyrme interactions~\cite{Paar09}. This method has the
advantage of consistency, however, it misses relevant correlations
which, as we will demonstrate in the present paper, are crucial to
derive stellar electron capture rates for neutron-rich nuclei.

The statistical way for the calculation of temperature-dependent GT
and first-forbidden strength functions was first applied
in~\cite{coo84} to study electron capture on neutron-rich nuclei.
The most advances realization of this way is presently performed in
the framework of the shell-model Monte-Carlo (SMMC)
method~\cite{koo97}.  It was found~\cite{rad97} that with increasing
temperature the GT centroids shift to lower excitation energies and
the widths of the distributions increase with the appearance of
low-lying strength. Both effects arise from thermally excited
states, i.e., the Brink hypothesis is not supported by SMMC
calculations.  In spite of its advantages, the SMMC method only
yields the lowest moments of the GT strength distributions which
introduce some inaccuracies into the rate calculations. Moreover,
the SMMC method has restrictions in its applicability to odd-odd and
odd-A nuclei at low temperatures.

Thus, the problem of an accurate description of the GT strength
distribution at finite temperatures and reliable estimates of
stellar weak-interaction rates is not solved completely yet: The
shell-model diagonalization approach allows for detailed
spectroscopy, but partially employs the Brink hypothesis. The SMMC
method is free from this disadvantage, but cannot provide a detailed
strength distribution. Moreover, present computer capabilities allow
the application of the shell-model diagonalization method only to
nuclei in the iron region ($A=45-65$), whereas the SMMC approach can
in principle be applied to all nuclei. However, such calculations
are rather time-consuming and have therefore been limited to about
200 nuclei with mass numbers $A=65-120$~\cite{lang03,juodagalvis09},
while weak processes in more massive and neutron-rich nuclei also
play an important role in various astrophysical scenarios.
Therefore, alternative methods to deal with Gamow-Teller strength
distributions and weak-interaction rates at finite temperatures are
desirable.

In this paper we study the temperature dependence of the
Gamow-Teller strength applying the proton-neutron quasiparticle
RPA~\cite{hal67} extended to finite temperature by the
Thermo-Field-Dynamics (TFD) formalism~\cite{ume75,ume82}. This
technique has the advantage that it does not rely on Brink's
hypothesis. The energies of the GT transitions and corresponding
transition strengths are calculated as functions of the nuclear
temperature. In this paper, we apply this method for the
calculations of weak-interaction rates on iron group nuclei and
neutron-rich nuclei beyond $pf$-shell. However, this method is not
only restricted to these nuclei but can be applied to heavier nuclei
as well. In addition,  it allows calculations of the strength
distributions for forbidden transitions which contribute
significantly to weak-interaction rates at high densities. Although,
in the present paper, we restrict our study to the one-phonon
approach, the method can be extended to higher phonon admixtures
thus yielding more detailed strength distributions.

The paper is organized as follows. In Sec.~\ref{formalism}, some
important features of the TFD formalism with application to the
nuclear structure problem at finite temperatures are presented.  The
finite temperature (TQRPA) equations, which describe the strength
distribution of charge-exchange transitions in hot nuclei, are given
in this section as well.  In Sec.~\ref{el_capt_rate} the necessary
formulae to calculate electron capture rates in a stellar
environment are introduced. Results for the GT strength
distributions and electron capture rates in $^{54,\,56}$Fe are
presented in Sec.~\ref{Fe}. Here, we also compare the results with
those from the shell-model diagonalization approach. In
Sec.~\ref{Ge}, we study the temperature dependence of GT and
first-forbidden strength distributions in the neutron-rich isotope
${}^{76}$Ge. The corresponding electron capture cross sections and
rates are calculated and compared with those obtained based on a
hybrid SMMC+RPA model~\cite{lan01_2}. Conclusions are drawn in
Sec.~\ref{conclusion}.

\section{Formalism\label{formalism}}

As a method to study a thermal behavior of quantum many-body systems
the TFD method has two attractive features: a) temperature effects
arise explicitly as $T$-dependent vertices, providing a convenient
starting point for various approximations; b) temperature and time
are independent variables. The first feature allows for
straightforward extensions of well-established zero-temperature
approximations. It has been employed previously
in~\cite{tana88,hats89,civi93,kos94,KVW97} where selected nuclear
structure problems at finite temperatures were considered.

The standard TFD formalism treats a many-particle system in thermal
equilibrium with a heat bath and a particle reservoir in the grand
canonical ensemble. The thermal average of a given operator $A$  is
calculated as the expectation value in a specially constructed,
temperature-dependent state $|0(T)\rangle$, which is termed the
thermal vacuum. This expectation value is equal to the usual grand
canonical average of $A$.

To construct the state $|0(T)\rangle$, a formal doubling of the
system degrees of freedom is introduced. In TFD, a tilde conjugate
operator~$\widetilde A$ -- acting in the independent Hilbert space
-- is associated with $A$, in accordance with properly formulated
tilde conjugation rules~\cite{ume75,ume82,oji81}. For a system
governed by the Hamiltonian~$H$ the whole Hilbert space is now
spanned by the direct product of the eigenstates of~$H$
(${H|n\rangle=E_n|n\rangle}$) and those of the tilde
Hamiltonian~$\widetilde H$, both having the same eigenvalues
(${\widetilde H|\widetilde n\rangle=E_n|\widetilde n\rangle}$). In
the doubled Hilbert space, the thermal vacuum is defined as the
zero-energy eigenstate of the so-called thermal Hamiltonian
${{\mathcal H}=H-\widetilde H}$  and it satisfies the thermal state
condition~\cite{ume75,ume82,oji81}
\begin{equation}\label{TSC}
A|0(T)\rangle = \sigma\,{\rm e}^{{\mathcal H}/2T} {\widetilde
A}^\dag|0(T)\rangle,
\end{equation}
where  $\sigma=1$ for bosonic~$A$ and $\sigma=i$ for fermionic $A$.

The important point is that in the doubled Hilbert space the
time-translation operator is not the initial Hamiltonian~$H$, but
the thermal Hamiltonian~${\mathcal H}$. This means that the
excitations of the thermal system are obtained by the
diagonalization of~${\cal H}$. As it follows from the definition of
$\mathcal H$ each of its eigenstates with positive energy has a
counterpart -- the tilde-conjugate eigenstate -- with negative
energy, but the same absolute value. This is a way to treat
excitation- and de-excitation processes at finite temperatures
within TFD.

Obviously, in most of practical cases one cannot diagonalize
$\mathcal H$ exactly. Usually, one resorts to certain approximations
such as the Hartree-Fock Bogoliubov mean field theory (HFB) and the
Random-Phase Approximation (RPA) (see e.g.~\cite{hats89}). In what
follows the formal part of the TFD studies for charge-exchange
excitations in hot nuclei is based in part on the results
of~\cite{dzh08,dzh09}.

In our present study we use a phenomenological nuclear Hamiltonian
consisting of a static mean field, BCS pairing interactions and
separable multipole and spin-multipole particle-hole interactions,
including isoscalar and isovector parts. This is usually referred to
as the quasiparticle-phonon model (QPM)~\cite{sol92}. It was used to
study the charge-exchange excitations in nuclei at zero temperature
in~\cite{kuz84,kuz85}. In principle, the QPM formalism enables one
to go beyond the QRPA and take into account the coupling of
quasiparticles and phonons. At finite temperatures this coupling was
considered in~\cite{kos94,dzh08}. However, in the present study we
restrict ourselves to the thermal QRPA.

The main line of the present discussion is very similar to the QPM
at $T=0$~\cite{sol92}. We begin with the thermal Hamiltonian, which
reads as
\begin{equation}\label{t_QPM}
{\cal H}_{\rm QPM} = H_{\rm QPM}-\widetilde H_{\rm QPM}={\cal
H}_{\rm sp}+{\cal H}_{\rm pair}+ {\cal H}_{\rm ph}~.
\end{equation}
The first step in the approximate diagonalization of ${\mathcal
H}_{\rm QPM}$ is the treatment of the pairing correlations. This is
done by two successive unitary transformations. The first is the
usual Bogoliubov $u, v$ transformation from the original particle
operators $a^\dag_{jm},~a^{\phantom{\dag}}_{jm}$ to the
quasiparticle ones $\alpha^\dag_{jm},~\alpha^{\phantom{\dag}}_{jm}$.
The same transformation is applied to tilde operators $\widetilde
a^\dag_{jm},\ \widetilde a^{\phantom\dag}_{jm}$, thus producing the
tilde quasiparticle operators $\widetilde\alpha^\dag_{jm},\
\widetilde\alpha^{\phantom\dag}_{jm}$.

The second transformation is the so-called thermal Bogoliubov
transformation~\cite{ume75,ume82}. It mixes the quasiparticle and
tilde quasiparticle operators, thus producing thermal quasiparticle
operators and their tilde partners:
$\beta^\dag_{jm},~\beta^{\phantom{\dag}}_{jm},~
\widetilde\beta^\dag_{jm},~\widetilde\beta^{\phantom{\dag}}_{jm}$.
We use  Ojima's~\cite{oji81} complex form of the thermal Bogoliubov
transformation
\begin{align}\label{T_tr}
  \beta^\dag_{jm}&=x_j\alpha^\dag_{jm}-i y_j\widetilde\alpha_{jm}~, \nonumber\\
  \widetilde\beta^\dag_{jm}&=x_j\widetilde\alpha^\dag_{jm}+i
  y_j\alpha_{jm}~,~~
  (x^2_j+y^2_j=1)~.
\end{align}
The reasons for this are given in~\cite{dzh08}.

The coefficients $u_j,\ v_j,\ x_j,\ y_j$  are found by
diagonalizing~${\cal H}_{\rm sp}+{\cal H}_{\rm pair}$ and demanding
that the vacuum of thermal quasiparticle obeys the thermal state
condition~\eqref{TSC}. This is equivalent to the minimization of the
thermodynamic potential for Bogoliubov quasiparticles. As a result
one obtains the following equations for $u_j,\ v_j$ and $x_j,\ y_j$:
\begin{eqnarray}
v_j & = & \frac{1}{\sqrt
2}\left(1-\frac{E_j-\lambda_{\tau}}{\varepsilon_j}\right)^{1/2},\
 u_j=(1-v_j^2)^{1/2}, \label{u&v} \\
y_j & = &
\left[1+\exp\left(\frac{\varepsilon_j}{T}\right)\right]^{-1/2},\
  x_j=\bigl(1-y^2_j\bigr)^{1/2}, \label{x&y}
\end{eqnarray}
where $\varepsilon_j=\sqrt{(E_j-\lambda_{\tau})^2+\Delta^2_{\tau}}$
and $\tau$ is the isospin quantum number $\tau = n, p $.

The pairing gap $\Delta_{\tau}$  and the chemical potential
$\lambda_\tau$ are the solutions to the finite-temperature BCS
equations
\begin{align}\label{BCS}
\Delta_\tau(T)&=\frac{G_\tau}{2}{\sum_j}^\tau(2j+1)(1-2y^2_j)u_jv_j,\nonumber\\
N_\tau&={\sum_j}^\tau(2j+1)(v^2_jx^2_j+u^2_jy^2_j),
\end{align}
where $N_\tau$ is the number of neutrons or protons in a nucleus and
${\sum}^\tau$ implies a summation over neutron or proton
single-particle states only. From the numerical solution of
Eqs.~(\ref{BCS}) it is found that the (pseudo)critical temperature
is $T_{\rm cr}\approx\frac{1}{2}\Delta_{\tau}(0)$ (see
e.g.~\cite{good81,civ83}) in accordance with the BCS theory.

With the coefficients $u_j,\,v_j$, $x_j,\,y_j$, defined by
Eqs.~(\ref{u&v},~\ref{x&y}), the one-body part of~${\cal H}_{\rm
sp}+{\cal H}_{\rm pair}$ reads
\begin{equation}\label{Ht_sp}
 {\cal H}_{\rm sp}+{\cal H}_{\rm
pair}\simeq\sum_\tau{\sum_{jm}}^\tau\varepsilon_j
(\beta^\dag_{jm}\beta^{\phantom{\dag}}_{jm}-\widetilde\beta^\dag_{jm}\widetilde\beta^{\phantom{\dag}}_{jm})
\end{equation}
and corresponds to a system of non-interacting thermal
quasiparticles.  The vacuum for thermal quasiparticles (hereafter
denoted by~${|0(T);{\rm qp}\rangle}$) is the thermal vacuum in the
BCS approximation. The states $\beta^\dag_{jm}|0(T);{\rm qp}\rangle$
have positive excitation energies whereas the corresponding
tilde-states $\widetilde\beta^\dag_{jm}|0(T);{\rm qp}\rangle$ have
negative energies.

The coefficients $y^2_j$ defined through~\eqref{x&y} determine the
average number of thermally excited Bogoliubov quasiparticles in the
BCS thermal vacuum
\begin{equation}
  \langle 0(T);{\rm qp}|
  \alpha^\dag_{jm}\alpha^{\phantom{\dag}}_{jm}
  |0(T);{\rm qp}\rangle=y^2_{j}
\end{equation}
and, thus, coincide with the thermal occupation factors of the
Fermi-Dirac statistics. Since the thermal vacuum $|0(T);{\rm
qp}\rangle$ contains a certain number of Bogoliubov quasiparticles,
excited states can be built on $|0(T);{\rm qp}\rangle$ by either
adding or removing a Bogoliubov quasiparticle. Because of
\begin{align}\label{meaning}
 \alpha^\dag_{jm}|0(T);{\rm qp}\rangle&=x_j\beta^\dag_{jm}|0(T);{\rm
 qp}\rangle,\notag\\
\alpha^{\phantom\dag}_{\overline{\jmath m}}|0(T);{\rm qp}
\rangle&=-iy_j\widetilde\beta^\dag_{\overline{\jmath m}}|0(T);{\rm
qp}\rangle\notag
\\
(\alpha^{\phantom\dag}_{\overline{\jmath
m}}=&(-1)^{j-m}\alpha^{\phantom\dag}_{j-m})
\end{align}
the first process corresponds to the creation of a non-tilde thermal
quasiparticle with positive energy, whereas the second process
creates a tilde quasiparticle with negative energy.

In the next step of the approximate diagonalization of ${\mathcal
H}_{\rm QPM}$, long-range correlations due to the particle-hole
interaction are taken into account within the proton-neutron TQRPA.
The part of~${\cal H}_{\rm ph}$ in~\eqref{t_QPM} responsible for
charge-exchange excitations reads
\begin{multline}\label{H_ph}
{\cal H}^{\rm ch}_{\rm ph}=
 -2\sum_{\lambda\mu}\kappa^{(\lambda)}_1(M^\dag_{\lambda\mu} M^{\phantom{\dag}}_{\lambda\mu}-
\widetilde M^\dag_{\lambda\mu}\widetilde
M^{\phantom{\dag}}_{\lambda\mu})\\
 -2\sum_{L\lambda\mu}\kappa^{(L\lambda)}_1( S^\dag_{L\lambda\mu} S^{\phantom{\dag}}_{L\lambda\mu}-
 \widetilde S^\dag_{L\lambda\mu}\widetilde S^{\phantom{\dag}}_{L\lambda\mu})~,
 \end{multline}
where $M^\dag_{\lambda\mu}$ and $S^\dag_{L\lambda\mu}$ are
single-particle multipole and spin-multipole operators:
\begin{align}\label{m&sm}
M^\dag_{\lambda\mu}&={\sum_{\genfrac{}{}{0pt}{1}{j_pm_p}{j_nm_n}}}
 \langle j_pm_p|i^\lambda r^\lambda Y_{\lambda\mu}(\theta,\phi)t^{(-)}|j_nm_n\rangle
 a^\dag_{j_pm_p}a^{\phantom{\dag}}_{j_nm_n},\notag
 \\
 S^\dag_{L\lambda\mu}&={\sum_{\genfrac{}{}{0pt}{1}{j_pm_p}{j_nm_n}}}
 \langle j_pm_p|i^\lambda r^\lambda[Y_L\,\sigma]^\lambda_\mu\,t^{(-)}|j_nm_n\rangle
 a^\dag_{j_pm_p}a^{\phantom{\dag}}_{j_nm_n},\notag
   \\
 &\bigl[Y_L\,\sigma\bigr]^\lambda_\mu=\sum_{M,\,m}\langle LM\,1m|\lambda\mu\rangle
 Y_{LM}(\theta,\phi)\sigma_m.
 \end{align}
The parameters $\kappa_1^{(\lambda)}$ and $\kappa_1^{(L\lambda)}$
denote the strength parameters of the isovector multipole and
spin-multipole forces, respectively. The states of natural parity
are generated by the multipole and spin-multipole~$L=\lambda$
interactions, while the spin-multipole interactions with
$L=\lambda\pm1$ are responsible for the states of unnatural parity.

Within the TFD formalism the TQRPA equations are derived in the
following way. First, ${\cal H}^{\rm ch}_{\rm ph}$ is written in
terms of the thermal quasiparticle operators. Then, the sum of
\eqref{Ht_sp} and ${\cal H}^{\rm ch}_{\rm ph}$ is diagonalized with
respect to charge-exchange thermal phonons.

The thermal phonon creation operator $Q^\dag_{\lambda\mu i}$ is
defined as a linear superposition of the proton-neutron thermal
two-quasiparticle operators
\begin{multline}\label{ch_phonon}
 Q^\dag_{\lambda\mu i}=\sum_{j_p j_n}
  \Bigl(
  \psi^{\lambda i}_{j_pj_n}[\beta^\dag_{j_p}\beta^\dag_{j_n}]^\lambda_\mu+
  \widetilde\psi^{\lambda i}_{j_pj_n}[\widetilde\beta^\dag_{\overline{\jmath_p}}
  \widetilde\beta^\dag_{\overline{\jmath_n}}]^\lambda_\mu\\+
  i\eta^{\lambda i}_{j_p j_n}[\beta^\dag_{j_p}
  \widetilde\beta^\dag_{\overline{\jmath_n}}]^\lambda_\mu+
  i\widetilde\eta^{\lambda i}_{j_pj_n}[\widetilde\beta^\dag_{\overline{\jmath_p}}
  \beta^\dag_{j_n}]^\lambda_\mu
  \\
  +
   \phi^{\lambda i}_{j_pj_n}[\beta_{\overline{\jmath_p}}\beta_{\overline{\jmath_n}}]^\lambda_{\mu}+
   \widetilde\phi^{\lambda i}_{j_p j_n}[\widetilde\beta_{j_p}\widetilde\beta_{j_n}]^\lambda_{\mu}\\+
   i\xi^{\lambda i}_{j_p j_n}[\beta_{\overline{\jmath_p}}
   \widetilde\beta_{j_n}]^\lambda_{\mu}+
   i\widetilde\xi^{\lambda i}_{j_pj_n}[\widetilde\beta_{j_p}
   \beta_{\overline{\jmath_n}}]^\lambda_{\mu}
   \Bigr),
\end{multline}
where $[~]^\lambda_\mu$ denotes the coupling of single-particle
angular momenta $j_n, j_p$ to total angular momentum $\lambda$. Now
the thermal equilibrium state is treated as the vacuum $|0(T);{\rm
ph}\rangle$ for the thermal phonon annihilation operators.

The thermal phonon operators are assumed to commute as bosonic
operators, i.e., ${[Q^{\phantom\dag}_{\lambda\mu
i},Q^{\dag}_{\lambda'\mu'
i'}]=\delta_{\lambda\lambda'}\delta_{\mu\mu'}\delta_{ii'}}$. This
assumption imposes the following constraint on the phonon
amplitudes:
 \begin{multline}\label{constraint}
 \sum_{j_p j_n}\Bigl(
  \psi^{\lambda i}_{j_pj_n}\psi^{\lambda i'}_{j_pj_n}+
 \widetilde\psi^{\lambda i}_{j_pj_n}\widetilde\psi^{\lambda i'}_{j_pj_n}+
  \eta^{\lambda i}_{j_pj_n}\eta^{\lambda i'}_{j_pj_n}\\+
  \widetilde\eta^{\lambda i}_{j_pj_n}\widetilde\eta^{\lambda i'}_{j_pj_n}
  -\phi^{\lambda i}_{j_pj_n}\phi^{\lambda i'}_{j_pj_n}-
  \widetilde\phi^{\lambda i}_{j_pj_n}\widetilde\phi^{\lambda i'}_{j_pj_n}\\-
   \xi^{\lambda i}_{j_pj_n} \xi^{\lambda i'}_{j_pj_n}-
  \widetilde\xi^{\lambda i}_{j_pj_n}\widetilde\xi^{\lambda i'}_{j_pj_n}\Bigr)=
      \delta_{ii'}.
      \end{multline}
Furthermore, the phonon amplitudes obey the closure relation.

Demanding that the vacuum of thermal phonons obeys the thermal state
condition~\eqref{TSC} and applying the variational principle to the
average value of thermal Hamiltonian with respect to one-phonon
states~${Q^\dag_{\lambda\mu i}|0(T);{\rm ph}\rangle}$ or
${\widetilde Q^\dag_{\overline{\lambda\mu i}}|0(T);{\rm
ph}\rangle}$\footnote{We take into account that the operator
$\widetilde Q^\dag_{\overline{\lambda\mu i}}=(-1)^{\lambda-\mu}
\widetilde Q^\dag_{\lambda-\mu i}$ transforms under spatial
rotations like a spherical tensor of rank~$\lambda$.} under the
constraints~\eqref{constraint} one gets a system of linear equations
for the amplitudes $\psi^{\lambda i}_{j_pj_n},\
\widetilde\psi^{\lambda i}_{j_pj_n},\ \eta^{\lambda i}_{j_pj_n},\
\widetilde\eta^{\lambda i}_{j_pj_n},$ etc. The system has a
nontrivial solution if the energy~$\omega_{\lambda i}$ of the
thermal one-phonon state obeys the following secular equation:
\begin{widetext}
\begin{equation}\label{sec_eq}
\left|
\begin{array}{cccc}
  X^{(+)}_{aa}-\dfrac{1}{\kappa_1^{(a)}} &  X^{(+-)}_{aa} & X^{(+)}_{ab} & X^{(+-)}_{ab}  \\
  X^{(-+)}_{aa}  & X^{(-)}_{aa}-\dfrac{1}{\kappa_1^{(a)}}  & X^{(-+)}_{ab} & X^{(-)}_{ab} \\
  X^{(+)}_{ab}   & X^{(+-)}_{ab} & X^{(+)}_{bb}-\dfrac{1}{\kappa_1^{(b)}} & X^{(+-)}_{bb} \\
  X^{(+-)}_{ab}  & X^{(-)}_{ab}  & X^{(-+)}_{ab} & X^{(-)}_{bb}-\dfrac{1}{\kappa_1^{(b)}} \\
\end{array}\right|=0,
\end{equation}
\end{widetext}
where $a\equiv\lambda$ and~${b\equiv\lambda\lambda}$ for excitations
of natural parity, while $a\equiv(\lambda-1)\lambda$
and~${b\equiv(\lambda+1)\lambda}$ for unnatural parity excitations.
The functions~$X^{(\pm)}_{cd}$, $ X^{(\pm\mp)}_{cd}$ (${c=a,\,b}$
and ${d=a,\,b}$) in~\eqref{sec_eq} are defined as
\begin{align}
X_{cd}^{(\pm)}(\omega)&=\frac{2}{\hat\lambda^2}\!\sum_{j_p j_n}
   f^{(c)}_{j_p j_n}f^{(d)}_{j_p j_n}\Biggl\{
      \frac{\varepsilon_{j_pj_n}^{(+)}(u^{(\pm)}_{j_pj_n})^2}
       {(\varepsilon_{j_pj_n}^{(+)})^2-\omega^2}
       \nonumber\\&\times(1\!-\!y^2_{j_p}\!-\!y^2_{j_n})-
      \frac{\varepsilon_{j_pj_n}^{(-)}(v^{(\mp)}_{j_pj_n})^2}
       {(\varepsilon_{j_pj_n}^{(-)})^2-\omega^2}(y^2_{j_p}\!-\!y^2_{j_n})\Biggr\},
\nonumber\\
   X_{cd}^{(\pm\mp)}(\omega)&=\frac{2\omega}{\hat\lambda^2}\sum_{j_p j_n}
  f^{(c)}_{j_p j_n}f^{(d)}_{j_p j_n}\Biggl\{
  \frac{u^{(\pm)}_{j_pj_n}u^{(\mp)}_{j_pj_n}}
       {(\varepsilon_{j_pj_n}^{(+)})^2-\omega^2}
       \nonumber\\\times(1-&y^2_{j_p}-y^2_{j_n})-
  \frac{v^{(\pm)}_{j_pj_n}v^{(\mp)}_{j_pj_n}}
       {(\varepsilon_{j_pj_n}^{(-)})^2-\omega^2}(y^2_{j_p}-y^2_{j_n})\Biggr\}.
\end{align}
Here $f^{(\lambda)}_{j_pj_n}$ and $f^{(\lambda L)}_{j_pj_n}$ denote
the reduced single-particle matrix elements of the multipole and
spin-multipole operators~\eqref{m&sm};
${u^{(\pm)}_{j_pj_n}=u_{j_p}v_{j_n}\pm v_{j_p}u_{j_n}}$ and
${v^{(\pm)}_{j_pj_n}=u_{j_p}u_{j_n}\pm v_{j_p}v_{j_n}}$;
${\varepsilon^{(\pm)}_{j_pj_n}=\varepsilon_{j_p}\pm\varepsilon_{j_n}}$;
$\hat\lambda=\sqrt{2\lambda+1}$.

Let us consider the secular equation in detail.  The poles
${\varepsilon^{(-)}_{j_pj_n}}$ which do not exist in the QRPA
equations at zero temperature arise from the crossed terms
$\beta^\dag\widetilde\beta^\dag$ in the thermal phonon operator
definition~\eqref{ch_phonon}. Due to these poles, new states appear
in a low-energy part of the thermal excitation spectrum. In contrast
to the zero temperature case, the negative solutions of the secular
equation now have a physical meaning. They correspond to the tilde
thermal one-phonon states and arise from
$\widetilde\beta^\dag\widetilde\beta^\dag$ terms in the thermal
phonon operator. As it was noted above, creation of a tilde thermal
quasiparticle corresponds to the annihilation of a thermally excited
Bogoliubov quasiparticle. Consequently, excitations of low- and
negative-energy thermal phonons correspond to transitions from
thermally excited nuclear states. Furthermore, when the pairing
correlations vanish (i.e., $T>T_{\rm cr}$), some poles no longer
contribute to the secular equation since the corresponding
numerators vanish. This is true for particle-particle and hole-hole
${\varepsilon^{(+)}_{j_pj_n}}$ poles as well as for particle-hole
${\varepsilon^{(-)}_{j_pj_n}}$ poles.

The expressions for the thermal charge-exchange phonon amplitudes
can be found in \cite{dzh09}. The amplitudes depend on both the
quasiparticle and the phonon thermal occupation numbers. Some
remarks are in order. TQRPA equations for GT excitations at finite
temperature were also derived in~\cite{civ01}. They were obtained
using the equation of motion method by replacing vacuum expectation
values by thermal averages, i.e., without applying the TFD formalism
and doubling the Hilbert space. Therefore, in contrast with the
present study negative solutions of the TQRPA equations were
neglected in~\cite{civ01}.

After diagonalization within the TQRPA the thermal Hamiltonian
$\mathcal H_{\rm QPM}$ becomes
\begin{equation}
{\cal H}_{\rm QPM}=\sum_{\lambda\mu i}\omega_{\lambda i}
   (Q^\dag_{\lambda\mu i}Q^{\phantom{\dag}}_{\lambda\mu i}
   -\widetilde Q^\dag_{\lambda\mu i}\widetilde Q^{\phantom{\dag}}_{\lambda\mu i}).
\end{equation}
The vacuum $|0(T);{\rm ph}\rangle$ of thermal phonons is the thermal
vacuum in the thermal quasiparticle RPA. Since we use the thermal
BCS approximation, which violates the particle number, the
charge-exchange thermal one-phonon states are superpositions of
states, which belong to the daughter nuclei $(N-1,Z+1)$ and
$(N+1,Z-1)$. They decouple at  temperatures $T \geq T_{\rm cr}$,
when the pairing correlations vanish. Then, if the state
${Q^\dag_{\lambda\mu i}|0(T);{\rm ph}\rangle}$ belongs to the
$(N\pm1,Z\mp1)$ nucleus, the state ${\widetilde
Q^\dag_{\overline{\lambda\mu i}}|0(T);{\rm ph}\rangle}$ is in the
$(N\mp1,Z\pm1)$ nucleus.

\section{Electron capture rates\label{el_capt_rate}}

Considering electron capture in stellar environments we make the
following assumptions: 1.~the temperature in the stellar interior is
so high that atoms are fully ionized, and the surrounding electron
gas is described by a Fermi-Dirac distribution, with temperature~$T$
and chemical potential~$\mu_e$. Neutrinos escape freely from the
interior of the star. Hence no Pauli blocking for neutrinos is
considered in the final state. 2.~the parent nucleus is in a thermal
equilibrium state treated as the thermal (phonon) vacuum.
3.~electron capture leads to charge-exchange transitions from the
thermal vacuum to thermal one-phonon states.

Under these circumstances the electron capture rate is the sum of
the transition rates from the thermal vacuum to the $i$-th thermal
one-phonon state of the multipolarity~$J$
\begin{equation}\label{rate}
\lambda^{\rm ec}=\frac{\ln 2}{6150\,{\rm sec}}\sum_J\sum_i
\Phi^{(+)}_{J i}F_{i}^{\,\rm ec} =\sum_J\sum_i\lambda^{\rm
ec}_{Ji}~.
\end{equation}
Here $\Phi^{(+)}_{J i}$ is the squared reduced matrix element of the
transition operator between the thermal phonon vacuum and a thermal
one-phonon state (see below); $F_{i}^{\,\rm ec}$ is a phase space
factor which depends on the transition energy~$E^{(+)}_{J i}$ and
can be found elsewhere~\cite{lan00}.

Denoting the proton-to-neutron ($p\to n$) transition operator with
multipolarity $J$ as ${D^{(+)}_{J}}$  one obtains the following
expression for the transition strength $\Phi^{(+)}_{J i}$
\begin{multline}\label{tr_nt}
\Phi^{(+)}_{J i}=
  \left|\langle 0(T);{\rm ph}\|Q_{JM i}D^{(+)}_J \|0(T);{\rm ph}\rangle\right|^2 \\=
\Bigl(\sum_{j_pj_n}(-1)^{j_n-j_p+J}
d^{(+)}_J(j_{p}j_{n})\Omega(j_pj_n;J i)\Bigr)^2,
\end{multline}
where~$d^{(+)}_J(j_{p}j_{n})=\langle j_{n}\|D^{(+)}_J\|j_{p}\rangle$
is a reduced single-particle matrix element of the transition
operator, and the function $\Omega(j_pj_n; J i)$ is given by
\begin{multline}\label{f1}
\Omega(j_pj_n;J i)=
   v_{j_p}u_{j_n}\bigl(x_{j_p}x_{j_n}\psi^{J i}_{j_pj_n}+
                       y_{j_p}y_{j_n}\widetilde\phi^{J
                       i}_{j_pj_n}\bigr)\\
                 +
   u_{j_p}v_{j_n}\bigl(y_{j_p}y_{j_n}\widetilde\psi^{J i}_{j_pj_n}+
                       x_{j_p}x_{j_n}\phi^{J i}_{j_p j_n}\bigr) \\
                -
   v_{j_p}v_{j_n}\bigl(x_{j_p}y_{j_n}\eta^{J i}_{j_pj_n}+
                       y_{j_p}x_{j_n}\widetilde\xi^{J
                       i}_{j_pj_n}\bigr)\\
     +
   u_{j_p}u_{j_n}\bigl(y_{j_p}x_{j_n}\widetilde\eta^{J i}_{j_pj_n}+
                        x_{j_p}y_{j_n}\xi^{J i}_{j_pj_n}\bigr).
\end{multline}
The transition strength to the tilde one-phonon state  can be easily
obtained from \eqref{tr_nt} and \eqref{f1} by changing non-tilde
phonon amplitudes by their tilde counterparts and vise versa. (The
expressions for the transition strengths $\Phi^{(-)}_{J i}$
corresponding to inverse $n\to p$ transitions are given
in~\cite{dzh09}. It has been proved in~\cite{dzh09} that the
approach used here fulfills the Ikeda sum rule for Fermi and
Gamow-Teller transitions.)

The transition energy (parent excitation energy) $E^{(+)}_{J i}$ can
be obtained from the energy shift between the proton subsystem of
the parent nucleus and the neutron subsystem of the daughter nucleus
including the proton-neutron mass difference. Thus we have
\begin{equation}\label{ph_tr_en}
E^{(+)}_{J i}=\omega_{J i}+(\Delta\mu_{np}+\Delta m_{np}),
\end{equation}
where $\Delta\mu_{np}=\mu_p-\mu_n$ is the difference between neutron
and proton chemical potentials and $\Delta m_{np}=m_n-m_p$ is the
neutron-proton mass splitting. Note, that at finite temperature the
energies $E^{(+)}_{J i}$ as well as $\omega_{J i}$ can be both
positive and negative. Thus, to the capture process thermal
one-phonon states with both positive and negative values of
$E^{(+)}_{J i}$ contribute to the rate.

In what follows in~\eqref{rate} we take into account the
contributions from allowed (Gamow-Teller and Fermi) transitions and
first-forbidden transitions.  The operators of  allowed Fermi and
Gamow-Teller transitions are taken in the standard form
\begin{eqnarray}\label{op_allowed}
D^{(+)}_{0^+}=g_Vt^{(+)}, \qquad D^{(+)}_{1^+} =
g_A{\boldsymbol\sigma}t^{(+)},
\end{eqnarray}
where $t^{(+)}$ is the isospin raising operator. For the operators
of the first-forbidden $n \to p$ transitions the non-relativistic
form is used
\begin{align}\label{op_forb}
D^{(+)}_{0^-}&= g_A\Bigl[\frac{\boldsymbol\sigma\cdot\boldsymbol
p}{m}+\frac{\alpha Z}{2R}i\boldsymbol\sigma\cdot\boldsymbol
r\Bigr]t^{(+)},
   \notag\\
D^{(+)}_{1^-}&=\Bigl[g_V\frac{\boldsymbol p}{m}-\frac{\alpha
Z}{2R}(g_A\boldsymbol\sigma\times\boldsymbol r - ig_V\boldsymbol
r)\Bigr]t^{(+)},
   \notag\\
D^{(+)}_{2^-}&=i\frac{g_A}{\sqrt
3}[\boldsymbol\sigma\cdot\boldsymbol r]^2_\mu
\sqrt{p^2_e+q^2_\nu}\,t^{(+)}.
\end{align}
In Eqs.~\eqref{op_allowed} and \eqref{op_forb} $\boldsymbol r,~
\boldsymbol p, \textrm{ and } \boldsymbol \sigma$ refer to the
coordinate, momentum and spin operators of a nucleon, $g_V=1$ and
$g_A=-1.25$ denote the vector and axial coupling constants, $\alpha$
is the fine structure constant, $Z$, $R$ are the charge and the
radius of the nucleus; $m$ is the nucleon mass and $p_e$ and $q_\nu$
denote the momenta of the incoming electron and outgoing neutrino,
respectively.

\section{\label{Fe} Iron isotopes}

\begin{figure}
\includegraphics[width=\columnwidth,trim=5 7
 11 10, clip]{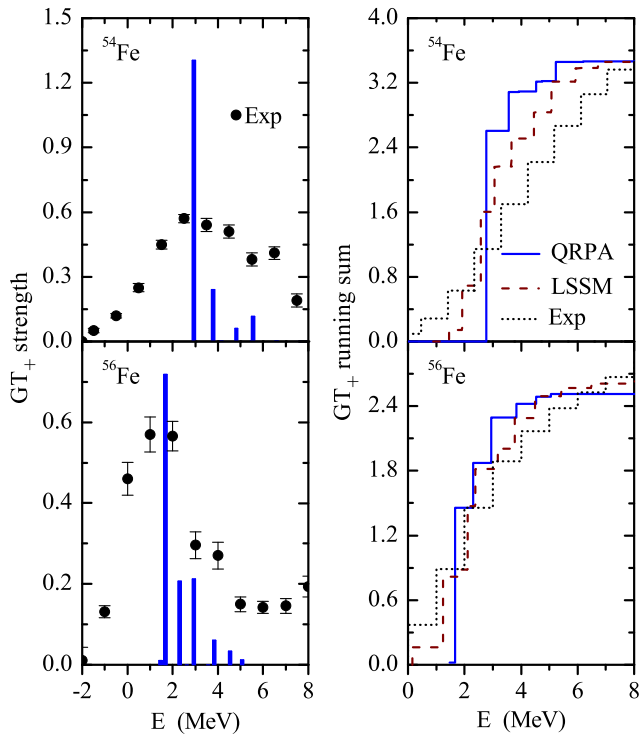}
\caption{(Color online) Left panels:  Comparison of GT$_+$
experimental  data \cite{ron93,kat94} with the calculated QRPA
strength distributions for  $^{54,56}$Fe. The QRPA peaks are scaled
by 0.5 for convenience. Right panels: Comparison of the GT$_+$
running sums corresponding to the experimental, QRPA, and LSSM
\cite{lan00} strength distributions. } \label{Fe_exp}
\end{figure}

In this section we discuss the numerical results for the $pf$-shell
nuclei~$^{54,56}$Fe. Experimental data are available for these
nuclei to test our calculations at zero temperature. Moreover, these
iron isotopes are among the most essential nuclei in their
importance for the electron capture process for the early
presupernova collapse~\cite{auf94,heg01}.

\begin{figure*}
\includegraphics[width=0.9\textwidth,trim=0 10
 0 0, clip]{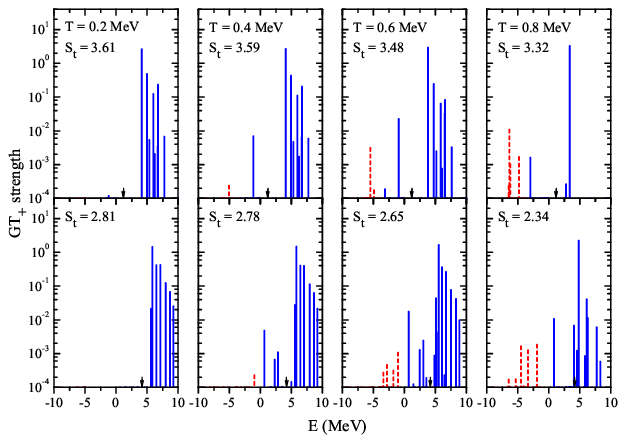}
\caption{(Color online) Temperature evolution of GT$_+$ strength
distributions for $^{54}$Fe (upper panels) and for $^{56}$Fe (lower
panels) versus parent excitation energy. The full (dashed) lines
refer to transitions to non-tilde (tilde) thermal one-phonon states.
$S_t$ is the total GT$_+$ strength. The arrows indicate the
zero-temperature threshold $Q=M_f-M_i$, where $M_{i,f}$ are the
masses of the parent and daughter nuclei. $Q({}^{54}{\rm
Fe})=1.21$~MeV and $Q({}^{56}{\rm Fe})=4.20$~MeV. }\label{Fe_GT}
\end{figure*}

The proton and neutron mean fields are described by spherically
symmetric Woods-Saxon potentials with parameters from \cite{che67}.
We only readjust the potential depths to fit the proton and neutron
binding energies of the parent nucleus to their experimental values.
The single-particle basis includes all discrete bound states as well
as selected quasi-bound states with large $j$ in the continuum. The
proton (neutron) pairing strength parameters $G_{p(n)}$ are fixed to
reproduce the odd-even mass difference through a four term
formula~\cite{pom97} involving the experimental binding
energies~\cite{audi93}.
 At $T=0$ the obtained proton and neutron  BCS energy
gaps are: ${\Delta_{p(n)}=1.52\,(0.0)}$\,MeV for
$^{54}$Fe,\footnote{$^{54}$Fe has a closed $1f_{7/2}$ neutron
subshell in our single-particle scheme.}
${\Delta_{p(n)}=1.57\,(1.36)}$\,MeV for $^{56}$Fe. The isovector
strength parameters  $\kappa_1^{(01)}$ and $\kappa_1^{(21)}$ are
adjusted to reproduce the experimental centroid energies of the
GT$_-$ and GT$_+$ resonances in the nuclei under
consideration~\cite{rap83,ron93,kat94}.  The corresponding values of
$\kappa_1^{(01)}$ and $k_1^{(21)}$ are in agreement with the rough
estimates in~\cite{cas76}. The spin-quadrupole interaction weakly
affects the GT strength distributions.

The total GT strengths calculated with the bare GT$_\pm$ operators
$\sigma t^{(\pm)}$ are ${S_+=6.}6$, $S_-=12.7$  in ${}^{54}$Fe and
$S_+=5.1$, $S_-=17.0$ in ${}^{56}$Fe respectively. These $S_{\mp}$
values obey the Ikeda sum rule (a small deviation is due to the
incompleteness of our single particle basis) but noticeably
overestimate experimental data (see e.g. \cite{rap83,ron93,kat94}).
This is common for any RPA (or QRPA) calculation of the GT strength
and is remedied by an effective value for the axial weak coupling
constant. We use $g^*_{A} = 0.74 g_{A}$ as in shell-model
calculations~\cite{lan00}.

In Fig.\ref{Fe_exp}, the experimental and theoretical (quenched)
distributions of GT$_{+}$ strengths are presented. Here we also
compare the GT$_+$ running sums corresponding to the experimental,
QRPA, and large scale shell-model (LSSM)~\cite{lan00} strength
distributions. One can see that the QRPA calculations reproduce the
resonance positions but not the fragmentation of the strength. It is
a well known fact that RPA calculations cannot describe the full
resonance width (at least in spherical nuclei) and produce only a
part of it, the so-called Landau width. The latter is quite small
for the GT resonance. As a result, the near threshold part of the
GT$_+$ strength, which corresponds to transitions to low-lying $1^+$
states in the daughter nuclei $^{54,56}$Mn, is not reproduced in our
calculations. In this respect the shell-model calculations are
clearly at an advantage.

\begin{figure*}
\includegraphics[width=0.9\textwidth,trim=0 10
 0 0, clip]{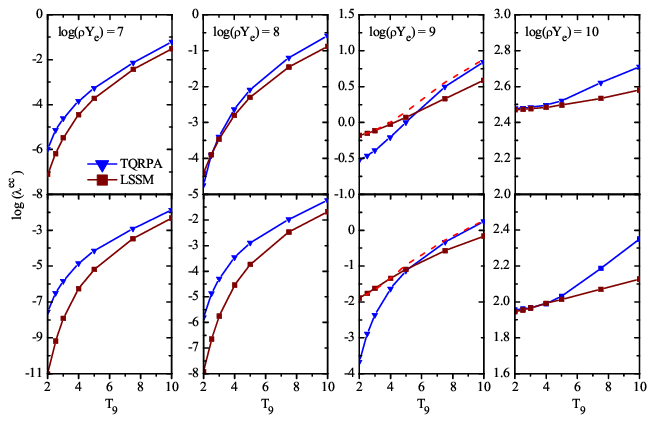}
\caption{(Color online) Electron capture rates for $^{54}$Fe (upper
panels) and $^{56}$Fe (lower panels) calculated using the  TQRPA
approach as a function of temperature ($T_9$ measures the
  temperature in $10^9$~K) and for selected values of density~$\rho
  Y_e$ (in g\,cm$^{-3}$). For comparison, the LSSM
  rates~\cite{lan01_1} are also shown. The dashed lines at $\log(\rho
  Y_e)=9$ correspond to the TQRPA rates calculated with the assigned
  near threshold strength (see text). }\label{Fe_rates}
\end{figure*}

We now turn to the temperature evolution of the GT$_+$ strength
distributions. The strength distributions for $^{54,56}$Fe at
several temperatures are shown in Fig.~\ref{Fe_GT}. All figures are
plotted as a function of the energy transfer $E$ to the parent
nucleus.

With increasing temperature, in our model, two effects occur that
influence the GT$_+$ strength distribution:

(i) At low temperatures, due to pairing, GT$_+$ transitions involve
the breaking of a proton Cooper pair associated with some energy
cost. This extra energy is removed at $T>T_{\rm cr}$ ($T_{\rm
  cr}\approx0.8$~MeV) and the peak in the GT$_+$ distribution moves to
smaller energies. Some extra energy has to be paid at low
temperatures to add one more nucleon to the neutron subsystem
of~$^{56}$Fe because of a non-zero neutron energy gap. Obviously,
this energy is also removed at~$T>T_{\rm cr}$.

(ii) GT$_+$ transitions, which are Pauli blocked at low temperatures
due to closed neutron subshells (e.g., $1f_{7/2}$ orbital), become
thermally unblocked with increasing temperature. Similarly, protons
which are thermally excited to higher orbitals can undergo GT$_+$
transitions. In TFD such transitions are taken into account
by~$\beta^\dag_{j_p}\widetilde\beta^\dag_{\overline{\jmath_n}}$,
$\widetilde\beta^\dag_{\overline{\jmath_p}}\beta^\dag_{j_n}$, and
$\widetilde\beta^\dag_{\overline{\jmath_p}}\widetilde\beta^\dag_{\overline{\jmath_n}}$
components of the thermal phonon. Because of thermally unblocked
transitions, some GT$_+$ strength appears well below the
zero-temperature threshold, including negative energies.

Due to the vanishing of the pairing correlations and appearance of
negative- and low-energy transitions, the centroids of the GT$_+$
strength distributions in $^{54,56}$Fe are shifted to lower
excitation energies at high temperatures. Our calculations indicate
that a temperature increase to 0.8~MeV results in the GT$_+$
centroid shifts of the order of $1.5$~MeV for $^{54}$Fe and
$2.5$~MeV for $^{56}$Fe. Thus the present approach violates Brink's
hypothesis. Similar results have been obtained in SMMC calculations
of the GT centroids at finite temperatures. We also observe (see
Fig.~\ref{Fe_GT}) a gradual decrease of the total GT$_+$ strength
when the temperature increases from zero to 0.8~MeV. Nevertheless,
as was pointed out above,  the present approach preserves the Ikeda
sum rule at finite temperatures.

The calculated GT$_+$ strength distributions have been used to
obtain the stellar electron capture rates for $^{54,56}$Fe. The
rates have been calculated for densities between $\log(\rho Y_e)=7$
and $\log(\rho Y_e)=10$ as a function of temperature~$T_9$
($T_9=10^9$~K and 1~MeV$\approx11.6~T_9$). The comparison between
the TQRPA rates and the large-scale shell-model
results~\cite{lan01_1} is presented in~Fig.~\ref{Fe_rates}.

As it must be, the electron capture rates increase with temperature
and density. Due to the larger value of the zero-temperature
threshold~$Q$ for $^{56}$Fe, both approaches yield a higher rate for
$^{54}$Fe than for $^{56}$Fe at a given temperature and density.
Both approaches give very similar values for the strength and the
location of the GT$_+$ resonance in $^{54,56}$Fe at $T=0$.
Therefore, the excellent agreement between the TQRPA and shell model
rates at $\log(\rho Y_e)=10$ and low temperatures ($\mu_e\approx
11$~MeV) is not surprising, since the rates are dominated by the
resonance contribution.

\begin{figure}
\includegraphics[width=\columnwidth,trim=10 10
 10 0, clip]{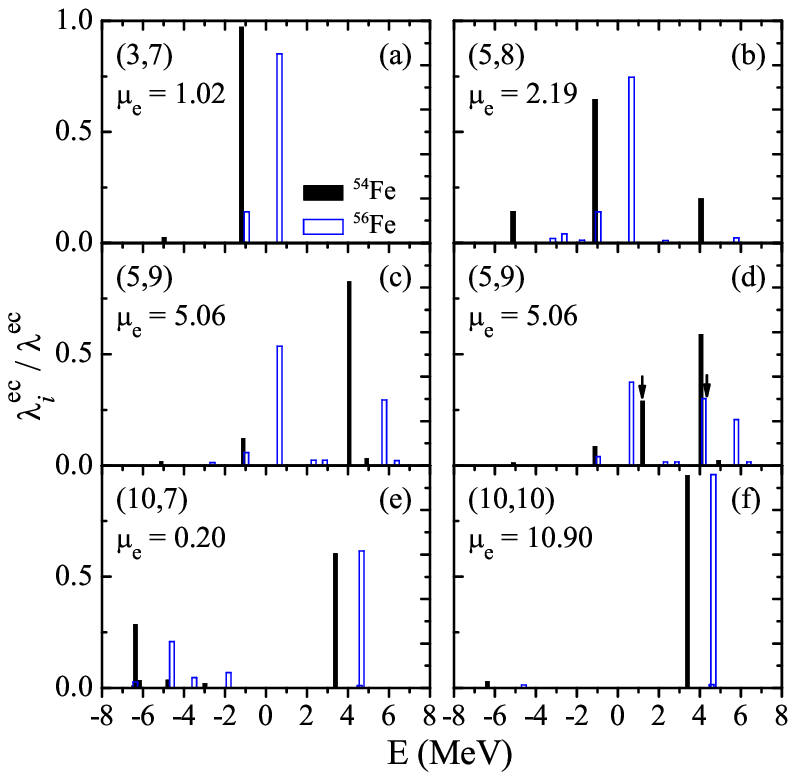}
\caption{(Color online) Relative contribution $\lambda^{\rm
ec}_{i}/\lambda^{\rm ec}$ of $i$-th thermal one-phonon state to the
electron capture rate on  $^{54,56}$Fe for selected values of
temperature and density, ($T_9,\log(\rho Y_e)$). The electron
chemical potential $\mu_{\rm e} $ is in units of MeV; $E$ is the
transition energy to $i$-th thermal one-phonon state. The arrows in
panel (d) indicate the relative contributions of the assigned near
threshold strengths (see text).
 }\label{rate_contr}
\end{figure}

The more interesting point is that at high temperatures the TQRPA
rates always surpass the shell-model ones. To understand this point,
it needs to be clarified which part of the TQRPA GT$_+$ strength
dominates the electron capture at a given temperature and density.
To this end we calculate the relative contributions $\lambda^{\rm
  ec}_{i}/\lambda^{\rm ec}$ of different thermal one-phonon states to
the capture rates for selected values of temperature and density,
($T_9,\log(\rho Y_e)$).  The results are depicted in
Fig.~\ref{rate_contr}.

At low temperatures and densities (Fig.~\ref{rate_contr}(a),(b)),
i.e., when $\mu_e$ is small and high-energy electrons from the tail
of the Fermi-Dirac distribution are not sufficiently available to
allow for efficient capture on the GT$_+$ resonance, the TQRPA
capture rates are dominated either by the negative-energy
($^{54}$Fe) or by the low-energy ($^{56}$Fe) part of the GT$_+$
strength which originates from thermally unblocked $p\to n$
transitions. The TQRPA rates are larger than those of the
shell-model at low $(T,~\rho)$ due to differences in the strength
and the energy of such transitions.  We note that in the shell model
evaluation negative-energy transitions were mainly included by
back-resonances; i.e.  by inverting the Fermi and GT$_-$ strength
distribution of $^{54}$Mn and $^{56}$Mn, respectively. Different to
the TQRPA approach the shell model GT$_-$ distributions of these
nuclei are highly fragmented due to correlations and have centroids
at rather high excitation energies in $^{54}$Fe and $^{56}$Fe which,
at low temperatures, are strongly suppressed by the Boltzmann
factor. In particular, the differences in energy positions of the
transitions are important since at low $(T,~\rho)$ the rates can
change drastically by a small change in a transition energy. To see
whether the TQRPA reliably predicts the energy and the strength of
negative- and low-energy transitions one needs to go beyond the
TQRPA.

At $\log(\rho Y_e)=9$ and $T_9<5$ ($\mu_e\approx 5.1$~MeV) the near
threshold part of the GT$_+$ strength dominates the capture rates.
Since this part is not reproduced within the TQRPA, the rates appear
to be smaller than the LSSM ones.  To test this hypothesis, the
capture rates at $\log(\rho Y_e)=9$ have been calculated -- guided
by the shell-model GT$_+$ distributions~\cite{lan00} -- assuming
that the near threshold GT$_+$ strengths for $^{54}$Fe and $^{56}$Fe
are 0.1 and 0.2, respectively. We therefore assign the value 0.1
(0.2) to the GT$_+$ strength in $^{54}$Fe ($^{56}$Fe) at the
zero-temperature threshold. (A similar method was used
in~\cite{fuller,auf94} to include the contribution of low-lying
transitions.) This yields a much better agreement between the TQRPA
and shell-model rates~(see Fig.~\ref{Fe_rates}). Thus, to improve
the reliability of the TQRPA for the calculation of stellar electron
capture rates at intermediate densities and low temperatures, the
fragmentation of the GT$_+$ resonance should be considered to
reproduce the near threshold GT$_+$ strength. At higher temperatures
the near threshold strength becomes less important.
In~Fig.~\ref{rate_contr}(c),(d) the relative contributions
$\lambda^{\rm ec}_{i}/\lambda^{\rm ec}$ at $(T_9,\log(\rho
Y_e))=(5,9)$ with and without the assigned near threshold strength
are depicted. As can be seen, the contribution from the near
threshold strength is not dominant.

When the temperature approaches $T_9\approx10$, the rates are
dominated by the strong transitions involving the GT$_+$ resonance
at low (Fig.~\ref{rate_contr}(e)) as well as at
high~(Fig.~\ref{rate_contr}(f)) densities. (Note that at $\log(\rho
Y_e)=7$ and  hight temperatures the contribution of negative-energy
transitions is non-negligible.) As was discussed above, the TQRPA
predicts that with increasing temperature the GT$_+$ resonance
shifts to lower excitation energies. This explains why the TQRPA
rates always surpass the LSSM ones at high temperatures.

Thus, only at high densities and low temperatures the TQRPA and LSSM
electron capture rates for $^{54,56}$Fe are in a good agreement. As
was mentioned above the disagreement at moderate densities and low
temperatures can be removed by considering the fragmentation of the
GT$_+$ strength. For a separable residual interaction used here this
can be done following the method developed within the
quasiparticle-phonon nuclear model, i.e., by taking into account the
phonon coupling. The interesting question is how the phonon coupling
affects the negative- and low-energy part of the GT$_+$ strength,
which dominates the capture rate at low temperatures and densities.
This is an open question and requires further investigations.

\section{Neutron-rich germanium isotopes\label{Ge}}

During  gravitational collapse the nuclear composition moves towards
higher mass number and more neutron-rich nuclei. Eventually nuclei
will have all neutron $pf$-shell orbits filled, with valence
neutrons in the $sdg$-shell (${N>40}$) and valence protons within
the $pf$-shell ($Z<40$). The Pauli principle blocks GT$_+$
transitions in such neutron-rich nuclei if the independent particle
model is used. It has been demonstrated in~\cite{coo84} that at high
enough temperatures, $T\sim1.5$~MeV, GT$_+$ transitions become
unblocked by thermal excitations which either move protons into
the~$1g_{9/2}$ orbital or remove neutrons from the~$pf$-shell. An
alternative unblocking mechanism, configuration mixing induced by
the residual interaction, was considered in~\cite{lan01_2}. Based on
this approach it was found  that electron capture on nuclei with
$N>40$ is also dominated by GT$_+$ transitions even at rather low
stellar temperatures, ${T\sim0.5}$~MeV. Contrary to~\cite{coo84}, it
was argued that unblocking effects due to mixing are not too
sensitive to increasing temperature.

Consistent calculations of the electron capture rates for
neutron-rich nuclei are not yet feasible in the shell model due to
the large model space. In~\cite{lan01_2} the capture rates have been
calculated adopting a hybrid model: The partial occupation numbers
calculated within the SMMC approach at finite temperature were used
in calculations based on the RPA. In this subsection, using the
germanium isotopes $^{76,78,80}$Ge as examples,  we apply the TQRPA
formalism to calculate electron capture rates on neutron-rich
nuclei. Particular attention is paid to the temperature dependence
of the unblocking effect.

For our TQRPA calculations the parameters of the model Hamiltonian
for $^{76,78,80}$Ge are chosen in the same manner as for
$^{54,56}$Fe. The sequence of single-particle levels obtained is
close to that used in~\cite{coo84} for $^{82}$Ge. For pairing gaps
we obtain: ${\Delta_{p(n)}=1.50~(1.57)}$~MeV for $^{76}$Ge,
${\Delta_{p(n)}=1.59~(1.42)}$~MeV for $^{78}$Ge, and
${\Delta_{p(n)}=1.39~(1.35)}$~MeV for $^{80}$Ge.  Our calculations
take into account both the allowed (GT and Fermi) and
first-forbidden transitions with $J\le2$. To generate  one-phonon
states of natural and unnatural parity we use the isovector
multipole and spin-multipole strength parameters
$\kappa_1^{(\lambda)}$, $\kappa_1^{(L\lambda)}$ ($\lambda=0,~1,~2$)
according to~\cite{bes75,cas76}

\begin{figure}
\includegraphics[width=\columnwidth,trim=0 8
 0 0, clip]{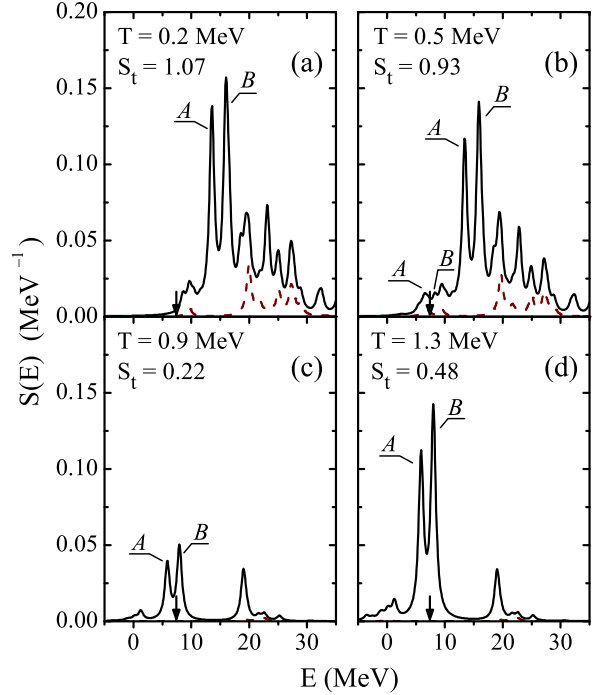}
\caption{(Color online) Strength distribution (folded) of allowed
($0^+$ and $1^+$) $p\to n$ transitions in~$^{76}$Ge at various
temperatures~$T$. $E$~denotes the transition energy. The
contribution of $0^+$ transitions is shown by the dashed line. $S_t$
is the total strength. The arrows indicate the zero temperature
threshold $Q(^{76}{\rm Ge})=M_f-M_i=7.52$~MeV. The letters label the
transitions: $A\equiv1f^p_{7/2}\to1f^n_{5/2}$,
$B\equiv1g^p_{9/2}\to1g^n_{7/2}$. }\label{76Ge_allowed}
\end{figure}

As a representative example, the strength distribution of allowed
$p\to n$ transitions from $^{76}$Ge for different values of
temperature is displayed in~Fig.~\ref{76Ge_allowed}. The
distributions have been folded with a Breit-Wigner function of~1~MeV
width. As it follows from our study as well as
from~\cite{coo84,lan01_2}, two single-particle transitions mainly
contribute to the total GT$_+$ strength in neutron-rich germanium
isotopes. These are the ${1g^p_{9/2}\to1g^n_{7/2}}$
particle-particle and ${1f^p_{7/2}\to1f^n_{5/2}}$ hole-hole
transitions. In an
  independent particle model both transitions are blocked at zero
  temperature. However, in the present model
  they become unblocked due to pairing
  correlations and thermal excitations.
Referring to Fig.~\ref{76Ge_allowed} it is observed that, with
increasing temperature, the peaks in the~GT$_+$ distribution shift
to lower excitation energies and the total strength decreases in the
vicinity of the critical temperature ($T_{\rm cr}\sim
0.8$~MeV)\footnote{The same effects were found in~\cite{dzh09} for
  $^{80}$Ge.}.  The shift is of the order of 8~MeV and, hence, cannot
be explained solely by removing the extra energy needed to break a
proton pair.

To explain both effects we neglect the residual particle-hole
interaction and consider the pairing interaction only. (As it
follows from our study, the position of the GT$_+$ peaks in
$^{76,78,80}$Ge is little affected by the inclusion of QRPA
correlations and thermal one-phonon states can be considered as
thermal two-quasiparticle states.) At finite temperature the GT$_+$
operator can excite configurations of four different types, namely,
$[\beta^\dag_{j_p}\beta^\dag_{j_n}]^1_\mu$,
$[\widetilde\beta^\dag_{\overline{\jmath_p}}\widetilde\beta^\dag_{\overline{\jmath_n}}]^1_\mu$,
$[\widetilde\beta^\dag_{j_p}\beta^\dag_{\overline{\jmath_n}}]^1_\mu$
, and
$[\widetilde\beta^\dag_{\overline{\jmath_p}}\beta^\dag_{j_n}]^1_\mu$.
The respective transition energies and transition strengths are
\begin{align}\label{tr&en}
E_1(j_p\to j_n)=&\varepsilon_{j_p}+\varepsilon_{j_n}+Q^*,\notag
\\
\Phi_1(j_p\to
j_n)&=(f^1_{j_pj_n})^2v^2_{j_p}u^2_{j_n}x^2_{j_p}x^2_{j_n};\notag
  \\
E_2(j_p\to j_n)=&-(\varepsilon_{j_p}+\varepsilon_{j_n})+Q^*,\notag
\\
\Phi_2(j_p\to
j_n)&=(f^1_{j_pj_n})^2u^2_{j_p}v^2_{j_n}y^2_{j_p}y^2_{j_n};\notag
  \\
E_3(j_p\to j_n)=&\varepsilon_{j_p}-\varepsilon_{j_n}+Q^*,\notag
 \\
\Phi_3(j_p\to
j_n)&=(f^1_{j_pj_n})^2v^2_{j_p}v^2_{j_n}x^2_{j_p}y^2_{j_n};\notag
 \\
E_4(j_p\to j_n)=&-(\varepsilon_{j_p}-\varepsilon_{j_n})+Q^*,\notag
 \\
\Phi_4(j_p\to
j_n)&=(f^1_{j_pj_n})^2u^2_{j_p}u^2_{j_n}y^2_{j_p}x^2_{j_n},
\end{align}
where $Q^*=\Delta\mu_{np}+\Delta m_{np}$ (see~Eq.~\eqref{ph_tr_en}).
In the following  $j_p\to j_n$ refer either to the
${1g^p_{9/2}\to1g^n_{7/2}}$ or the ${1f^p_{7/2}\to1f^n_{5/2}}$
transition.

At $T<T_{\rm cr}$ the excitation of the
$[\beta^\dag_{j_p}\beta^\dag_{j_n}]^1_\mu$ configuration dominates
the strength distribution because of the factor
$x_{j_p}^2x_{j_n}^2\sim1$ in $\Phi_1(j_p\to j_n)$. Therefore, at
relatively low temperatures, when configuration mixing induced by
pairing correlations in the ground state is the main unblocking
mechanism, the position of the GT$_+$ peaks is given by
$E_1(1g^p_{9/2}\to1g^n_{7/2})$ and $E_1(1f^p_{7/2}\to1f^n_{5/2})$
(Fig.~\ref{76Ge_allowed}(a),(b)). With increasing temperature states
having internal configurations other than those of the nuclear
ground state gain statistical weight and, in particular, the pairing
correlations in these excited states decrease. When the pairing
correlations disappear and the factors $v^2_{j_p}u^2_{j_n}$ in
$\Phi_1(j_p\to j_n)$ become zero, the peaks considered completely
vanish (Fig.~\ref{76Ge_allowed}(c)). The value of $\Phi_2(j_p\to
j_n)$ becomes zero as well.  At $T\ge T_{\rm cr}$   the poles
$\varepsilon_{1g^p_{9/2}}+\varepsilon_{1g^n_{7/2}}$  and
$\varepsilon_{1f^p_{7/2}}+\varepsilon_{1f^n_{5/2}}$ no longer
contribute to the secular equation~\eqref{sec_eq}.

\begin{figure}
\includegraphics[width=\columnwidth,trim=15 13
 10 13, clip]{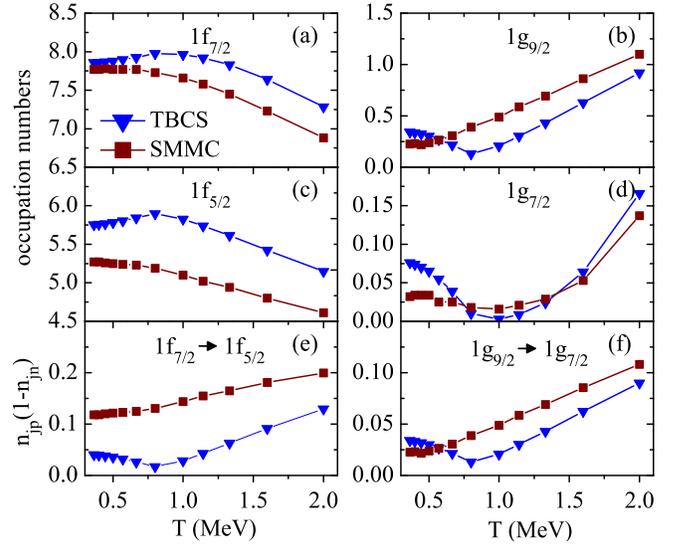}
\caption{(Color online) Upper panels: Occupation numbers for the
$1f_{7/2}$, $1g_{9/2}$ proton orbitals in $^{76}$Ge as a function of
temperature. Middle panels:  Occupation numbers for the $1f_{5/2}$,
$1g_{7/2}$ neutron orbitals.   Lower panels: The unblocking
probabilities for the $1f^p_{7/2}\to 1f^n_{5/2}$ and $1g^p_{9/2}\to
1g^n_{7/2}$ transitions.}\label{unblocking}
\end{figure}

At $T>T_{\rm cr}$ GT$_+$ transitions from (to)  thermally occupied
(unblocked) orbitals dominate the strength distribution. Such
transitions correspond to the excitation of the
$[\widetilde\beta^\dag_{{1g^p_{9/2}}}\beta^\dag_{1g^n_{7/2}}]^1_\mu$
and
$[\beta^\dag_{1f^p_{7/2}}\widetilde\beta^\dag_{{1f^n_{5/2}}}]^1_\mu$
configurations and their energies  are
$E_4(1g^p_{9/2}\to1g^n_{7/2})$ and $E_3(1f^p_{7/2}\to1f^n_{5/2})$,
respectively. Neglecting $\Delta m_{np}$, these energies are the
energy difference between the final and initial single-particle
states, i.e., $E_4(1g^p_{9/2}\to1g^n_{7/2})\approx
E_{1g^n_{7/2}}-E_{1g^p_{9/2}}$ and
$E_3(1f^p_{7/2}\to1f^n_{5/2})\approx E_{1f^n_{5/2}}-E_{1f^p_{7/2}}$.
Because of the thermally unblocked transitions the  GT$_+$ peaks
appear near the zero temperature
threshold~(Fig.~\ref{76Ge_allowed}(d)).

Thus, in contrast to~\cite{lan01_2}, we find that the unblocking
effect for GT$_+$ transitions in neutron-rich nuclei is sensitive to
increasing temperature. No shift to lower excitation energies for
the GT$_+$ peaks or decrease of the total GT$_+$ strength in the
vicinity of the critical temperature were observed
in~\cite{lan01_2}.  To understand these differences we compare again
the approximations underlying the present model and the hybrid
approach of \cite{lan01_2}. The TQRPA has the virtue of consistency.
It describes correlations by configuration mixing derived from a
pairing interaction up to the 2p2h level. In the hybrid model
occupation numbers at finite temperature are calculated within the
SMMC approach, accounting for all many-body npnh correlations
induced by a pairing+quadrupole residual interaction. These
occupation numbers have then been used to define a thermal ground
state  which is the basis of an RPA approach to calculate the
capture cross sections, considering only 1p1h excitations on the top
of this ground state. Therefore the hybrid model does not include
explicitly 2p2h pairing correlations when calculating strength
distributions.

Repeating our observation from above, the TQRPA has two distinct
transitions to overcome Pauli blocking. Using, for the sake of
simplicity the language of the Independent Particle Model, in the
TQRPA GT transitions can occur from a configuration-mixed states
with 0p0h and 2p2h components. These transitions lead to excited
states with centroids which are shifted by the excitation energy of
two particles which are raised across the $pf$--$sdg$ shell gap,
which corresponds to about 8 MeV for $^{76}$Ge. The two peaks
observed in the TQRPA GT strength distribution in
Fig.~\ref{76Ge_allowed}(a),(b) correspond to these two transitions.
As the 2p2h component has two neutron holes, GT transitions into
these holes are not Pauli blocked.  Hence these transitions are
relatively strong within the TQRPA model at low temperatures.  On
the other hand, GT transitions between pure 0p0h components are
Pauli blocked. Transitions to final states corresponding to the
lower centroid are only possible due to the small mixing of 2p2h
configurations into the final states. Hence the GT strength
corresponding to the lower peak is rather weak at low
temperatures~(Fig.~\ref{76Ge_allowed}(b)). The relative weight of
the transition strength between these two peaks changes with
increasing temperature due to the growing of the thermal excitations
and the decreasing correlations induced by pairing. The later effect
dominates at modest temperatures. As a consequence, the strength in
the upper peak decreases, while the lower peak grows and, at
temperatures beyond the critical temperature $T_{\rm cr}$, dominate
the GT strength distribution in the TQRPA model
(Fig.~\ref{76Ge_allowed}(d)).

 It is found that in the SMMC approach  many-body correlations lead to
much stronger excitation of particles (mainly neutrons) across the
$pf$--$sdg$ shell gap than found in the TQRPA model. This is
demonstrated in Fig.~\ref{unblocking}(a)-(d), which compares the
SMMC and thermal BCS occupation numbers for various orbitals as
function of temperature. While BCS predicts only 0.3 neutrons to be
excited out of the $pf$-shell at $T=0$ for $^{76}$Ge, this number is
about 1.2 for the SMMC, which is actually also smaller than the
number of neutron excitations across the shell gap, derived recently
experimentally ($2.48\pm0.30$)~\cite{ge76occ}. Obviously the
differences in occupation numbers leads to a larger Pauli unblocking
within the SMMC approach than found in TQRPA model. Correspondingly
the RPA calculation on top of the SMMC occupation numbers predicts
more GT strength in the energy range around 10 MeV (corresponding to
the 0p0h centroid in the TQRPA calculation) which, as we will show
below, results also in higher capture cross sections at low
temperatures. In passing we note that, at low temperatures, the SMMC
predicts a larger number of neutron excitations, but a smaller
number of proton excitations across the $pf$--$sdg$ shell gap, then
the TQRPA model. This underlines the importance of pn-correlations,
induced by isovector pairing and quadrupole interactions in the SMMC
approach. Such pn-correlations are not considered in the present
TQRPA calculations.

In the SMMC the many-body correlations induced by the
pairing+quadrupole interaction also yield a significantly smaller
temperature dependence in the occupation numbers than observed in
the TQRPA approach. In this model the energies of the unblocked
GT$_+$ transitions essentially depend on temperature: at $T>T_{\rm
cr}$, when unblocking is due to thermal excitations (thermal
unblocking), they are smaller than the ones at $T<T_{\rm cr}$, when
unblocking is due to configuration mixing. Obviously, the
significant shift of the GT$_+$ peaks  to lower excitation energies
favors electron capture. One can conclude that at $T>T_{\rm cr}$
GT$_+$ transitions in neutron-rich nuclei are more unblocked than at
$T<T_{\rm cr}$.

To explain the second effect, we consider the total strength for the
$j_p\to j_n$ transition
\begin{equation}
S_t(j_p\to j_n)=\sum_i \Phi_i(j_p\to
j_n)=(f^1_{j_pj_n})^2n_{j_p}(1-n_{j_n}),
\end{equation}
where $n_{j_\tau}$ is the proton ($\tau=p$) or neutron ($\tau=n$)
occupation factor
\begin{equation}\label{occup_prob}
n_{j}=\langle 0(T);{\rm qp}|
  a^\dag_{jm}a^{\phantom{\dag}}_{jm}
  |0(T);{\rm qp}\rangle=u^2_jy^2_{j}+v^2_jx^2_{j}.
\end{equation}
The value of $n_{j_p}(1-n_{j_n})$ determines the unblocking
probability for the $j_p\to j_n$ transition. As it follows from
Eq.~\eqref{occup_prob} the unblocking probability depends on the
temperature through the coefficients of both the Bogoliubov and
thermal transformation, i.e., it is determined by both configuration
mixing and thermal excitations. We note again that at~$0<T<T_{\rm
cr}$ the total strength~$S_t$ of the unblocked particle-particle or
hole-hole transition is not concentrated in only one peak, but as it
follows from~\eqref{tr&en}, is fragmented into four parts.
Fig.~\ref{unblocking}(e),(f) show the unblocking probabilities for
the ${1f^p_{7/2}\to1f^n_{5/2}}$ and ${1g^p_{9/2}\to1g^n_{7/2}}$
transitions as a function of temperature.

\begin{figure}
\includegraphics[width=\columnwidth, trim=0 6
 0 0, clip]{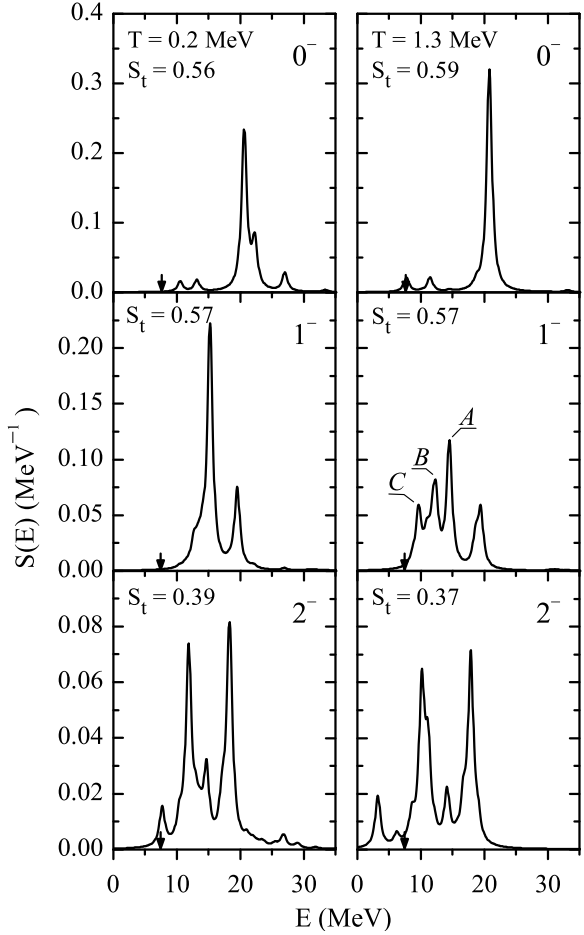}
\caption{Folded strength distributions  of first forbidden
$0^-,~1^-$ and $2^-$ $p\to n$ transitions in~$^{76}$Ge at
$T=0.2$~MeV (left panels) and~$T=1.3$~MeV (right panels); $E$~is the
transition energy. The strength distributions for  the $2^-$
multipole correspond to 25~MeV electrons. $S_t$ is the total
strength. The arrows indicate the zero temperature threshold. The
letters label the $1^-$ transitions:
$A\equiv1f^p_{7/2}\to2d^n_{5/2}$, $B\equiv1f^p_{5/2}\to1g^n_{7/2}$,
$C\equiv1f^p_{7/2}\to1g^n_{9/2}$.}\label{76Ge_forb}
\end{figure}

As seen from the figure the unblocking probabilities for both
transitions have a minimum at the critical temperature. It is
apparent that this minimum occurs because at $T_{\rm cr}$ pairing
correlations vanish while thermal effects are not yet sufficiently
strong to occupy the $1g_{9/2}$ proton orbit or unblock the
$1f_{5/2}$ neutron orbit. As a result the total transition strength
$S_t$ decreases in the vicinity of the critical temperature. In
contrast, this minimum is absent in the SMMC unblocking
probabilities (see Fig.~\ref{unblocking}(e),(f)). Here the residual
interaction introduces a slight, but gradual increase of the
probability with temperature. At $T > 1.5$ MeV the SMMC and TQRPA
results converge as is expected in the high temperature limit.

The fact that crossing shell gaps by correlations is a rather slowly
converging process which requires the consideration of
multi-particle-multi-hole configurations has already been observed
in large-scale diagonalization shell model calculations, e.g.
studying the calcium isotope shifts \cite{Caurier} or the M1
strength in argon isotopes \cite{Lisetzki}.

For neutron-rich nuclei the contribution of first forbidden $p\to n$
transitions to electron capture is not
negligible~\cite{coo84,lan01_2}.  The strength distributions of
first-forbidden $0^-,~1^-$, and $2^-$ transitions in $^{76}$Ge are
shown in Fig.~\ref{76Ge_forb} for temperatures
${T=0.2}$~and~$1.3$~MeV. The distributions have been folded by the
same procedure used above for the allowed transitions. As is seen
from the figure, a temperature increase weakly affects the peaks in
the $0^-,~2^-$ strength distributions. The reason is that these are
dominated by particle-hole transitions whose energy depends only
weakly on temperature (in contrast to particle-particle and
hole-hole transitions). With increasing temperature the peaks
slightly shift to lower excitation energies due to the vanishing of
the pairing correlations and some transition strength appears below
the zero temperature threshold due to thermally unblocked
transitions.

Finite temperature induces a significant spread in the $1^{-}$
transition strength distribution.  The spread can be easily
explained. At $T=0.2$~MeV the main peak in the distribution is
generated by three single-particle transitions:
$1f^p_{7/2}\to2d^n_{5/2}$, $1f^p_{5/2}\to1g^n_{7/2}$, and
$1f^p_{7/2}\to1g^n_{9/2}$. The first is a particle-hole transition
and its energy depends only slightly on temperature. The second and
third ones are particle-particle and hole-hole transitions,
respectively. As discussed above, the energies of particle-particle
and hole-hole transitions at $T>T_{\rm cr}$ are noticeably lower
than that at $T<T_{\rm cr}$. Therefore, at $T=1.3$~MeV, the peak is
fragmented into three parts, resulting in a broadening of the $1^-$
strength distribution. The $1^-$ peak at $E=19$~MeV is generated by
the particle-hole transition $1f^p_{7/2}\to1g^n_{7/2}$ and, hence,
its position and strength almost do not depend on the temperature.

To reveal the importance of thermal unblocking for GT$_+$
transitions in neutron-rich nuclei we perform electron capture cross
sections calculations. Within the present approach, the total cross
section for capture of an electron with energy $E_e$ on a nucleus
with charge~$Z$ is given by
\begin{equation}\label{EcCrSect}
\sigma(E_e,
T)=\frac{G^2_w}{2\pi}F(Z,E_e)\sum_{Ji}(E_e-E^{(+)}_{Ji})^2\Phi^{(+)}_{Ji},
\end{equation}
where $G_w$ is the weak interaction coupling constant; $F(Z,E_e)$ is
the Fermi function that accounts for the Coulomb distortion of the
electron wave function near the nucleus (see, for,
example,~\cite{lan00}). Only allowed and first forbidden transitions
are involved in the sum over~$J$ in the present study.

\begin{figure}
\includegraphics[width=\columnwidth,trim=19 19
 22 0, clip]{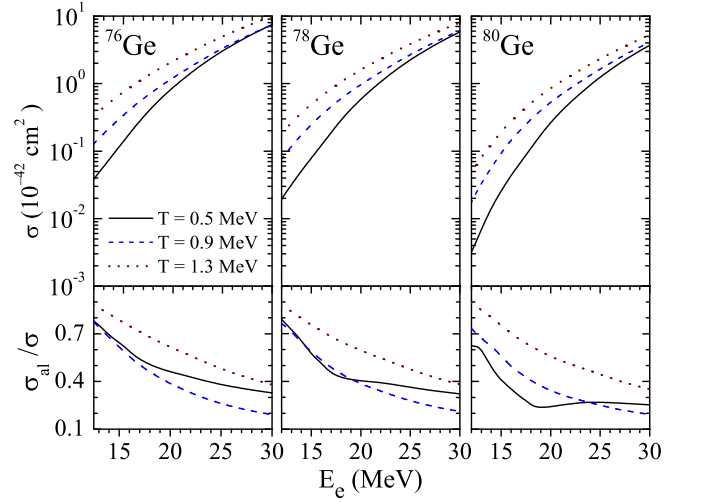}
\caption{(Color online) Electron capture cross sections (upper
panels) for~$^{76,80,80}$Ge calculated within the TQRPA approach for
various temperatures.  Relative contributions of allowed transitions
to the electron capture cross sections are shown in the lower
panels.}\label{cr_sect}
\end{figure}

\begin{figure*}
\includegraphics[width=0.9\textwidth, trim=0 10
 0 0, clip]{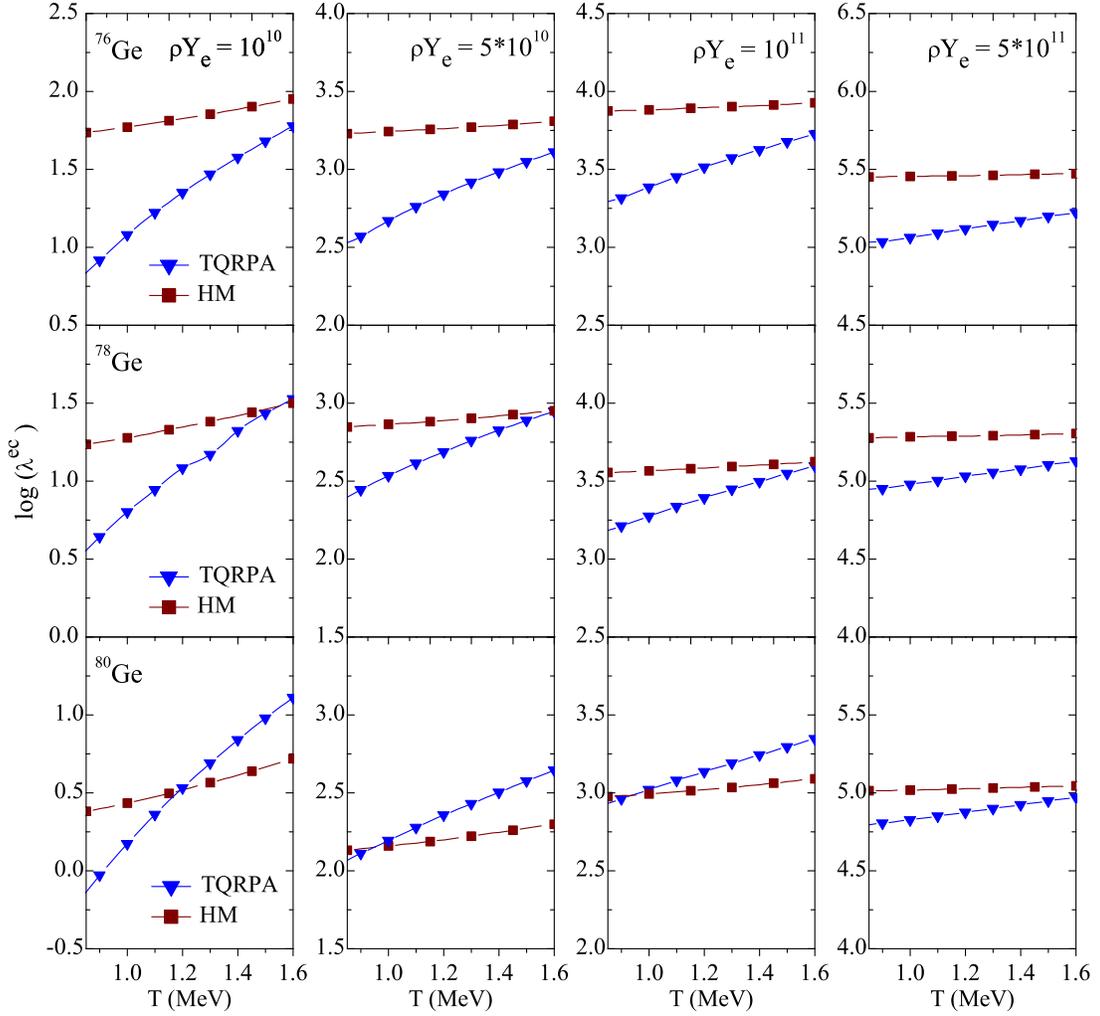}
\caption {(Color online) Electron capture rates for $^{76,78,80}$Ge
calculated using the TQRPA approach as a function of temperature and
for selected values of density~$\rho Y_e$ (in g\,cm$^{-3}$). For
comparison, the rates obtained by the hybrid model  are also
shown.}\label{Ge_rates}
\end{figure*}

In Fig.~\ref{cr_sect} the electron capture cross sections
for~$^{76,78,80}$Ge are shown for three different temperatures. The
temperature dependence of the cross sections is most pronounced at
moderate electron energies (${E_e\leq15}$~MeV): for $E_e=15$~MeV a
temperature increase from 0.5~MeV to 1.3~MeV results in an
enhancement of the cross sections by an order of magnitude. No such
enhancement was found in~\cite{lan01_2} (see below). To make clear
the reason of this enhancement, we calculate the relative
contribution of allowed transitions to the electron capture cross
sections. The results are displayed in the lower part of
Fig.~\ref{cr_sect}.

It is seen that at $E_e\leq15$~MeV electron capture is mainly
mediated by allowed transitions. Consequently the cross section
enhancement is caused by thermal  unblocking of GT$_+$ transitions
(the Fermi contribution to the cross sections is negligible).
Furthermore, because of thermal unblocking, the electron energy
below which electron capture is dominated by allowed transitions,
shifts to higher values: at $T=0.5$ and~$0.9$~MeV this energy is
$16-18$~MeV, while at $T=1.3$~MeV it is about~$25$~MeV. For larger
electron energies, the first forbidden transitions become
increasingly important. As the strength of the first forbidden
transitions is less sensitive to temperature change, the capture
cross sections at $E_e\sim30$~MeV depend only weakly on temperature.

The strong temperature sensitivity of the cross sections at low
electron energies reflects the temperature dependence of the TQRPA
GT strength distribution, as discussed above. This has mainly two
reasons. First, at low temperatures the dominant GT strength resides
at larger excitation energies than in the hybrid model. Second, in
the TQRPA the GT centroid shifts by several MeV, which is not
observed in the hybrid model. Both facts, amplified by the strong
phase space energy dependence, lead to a much stronger dependence of
the cross section in the TQRPA model than in the hybrid model. As
the GT contribution to the cross sections is larger in the hybrid
model than in the present TQRPA calculation, allowed transitions
dominate to higher electron energies in the former.

Fig.~\ref{Ge_rates}   compares the capture rates for $^{76,78,80}$Ge
as obtained in the hybrid and the TQRPA models. We note that the
hybrid model rates are noticeably larger than the present ones at
low temperatures. This is due to the increased unblocking
probability in the hybrid model caused by many-body correlations
which lead to a larger GT strength at lower excitation energies than
in the TQRPA approach. With increasing temperature and density the
differences between the rates of the two models become smaller. This
has two reasons. At first, with increasing temperature and density
the average electron energy increases and the rates become less
sensitive to details of the GT strength distribution. Secondly, the
GT strength distributions as calculated in the two models become
more similar with increasing temperature as discussed above.

We note that the rates obtained in both models for the temperature
and density regime, in which neutron-rich nuclei like those studied
here dominate the composition during supernova core collapse
($T>1$~MeV, {$\rho>5\times10^{10}$~g/cm$^3$}), are large enough so
that electron capture on nuclei dominates over capture on free
protons as has been predicted in \cite{lan01_2}.

\section{Conclusion\label{conclusion}}

In this work we have considered the case of GT$_+$ and
first-forbidden transitions in hot nuclei. We have applied the
proton-neutron quasiparticle RPA  extended to finite temperature by
the Thermo-Field-Dynamics formalism. The presented approach allows
to treat charge-exchange transitions in nuclei at finite temperature
without applying Brink's hypothesis. It fulfills the Ikeda sum rule
at finite temperature. As an example, we have calculated the
strength distribution for GT$_+$ transitions in $^{54,56}$Fe. We
have observed the downward shift of the GT$_+$ strength with
increasing temperature. This shift is caused by vanishing of pairing
correlations and the appearance of negative- and low-energy
transitions.  The shift of the GT$_+$ strength results in more
enhanced electron capture rates at high temperatures as compared to
those obtained from shell-model calculations. We have found that the
contribution of negative- and low-energy transition, i.e.,
transitions from thermally excited nuclear states, to electron
capture is non-negligible even at low temperatures. We have also
calculated the GT$_+$ strength distribution in the neutron-rich
$^{76,78,80}$Ge nuclei. It was found that the temperature increase
leads to a considerable (of the order of 8~MeV) downward shift of
the peaks in the strength distribution and reduces the total
transition strength in the vicinity of the critical temperature.
This makes the unblocking effect for neutron-rich nuclei quite
sensitive to increasing temperature in our model which is clearly
observed in the electron capture cross sections and rates for
$^{76,78,80}$Ge.

If we compare our results to those obtained within the
diagonalization shell model for the iron isotopes and within the
hybrid SMMC+RPA model for the neutron-rich germanium isotopes, the
importance of many-body correlations beyond those induced by pairing
in our TQRPA model become apparent. For the isotopes $^{54,56}$Fe
Pauli unblocking is unimportant as GT transitions are possible even
in the Independent Particle Model without correlations. The TQRPA
describes the centroid of the GT strength rather well. However, it
misses the low-lying GT strength, which is induced by multi-nucleon
correlations. Such low-lying strength in $^{54,56}$Fe, which is
experimentally observed and reproduced by the shell model, is
important for calculations of electron capture rates at low
temperatures. Its neglect leads to underestimation of the rates.

Pauli unblocking is crucial for the calculation of the GT strength
and the associated electron captures rates for the neutron-rich
germanium isotopes. Previous diagonalization shell model
calculations have shown that such cross-shell effects are rather
slowly converging with increasing correlations across the shell gap
and require the consideration of many-nucleon correlations. This is
in line with the observation that the SMMC approach, which accounts
for many-body configuration mixing, recovers significantly more
excitations across the $pf$--$sdg$ shell gap than is found within
the TQRPA approach, which, at low temperatures, derives Pauli
unblocking mainly from 2p2h pairing correlations. These differences
reflect themselves in different GT distributions and capture rates
at low temperatures and densities. However, the two models predict
rather similar capture rates for the collapse conditions at which
Pauli unblocking matters, making the capture on nuclei dominate over
capture on free protons.

In summary, we have presented here a method which allows to
calculate stellar electron capture rates at finite temperature in a
thermodynamically consistent way. This virtue makes it conceptually
superior to the hybrid approach of SMMC+RPA which has previously
been used to estimate such rates for neutron-rich nuclei. In the
present application of the model we have done a first step towards
its complete realization describing correlations within the TQRPA.
While this already recovers much of the essential physics, the
detailed comparison to shell model results implies that the model
must be extended to include correlations beyond TQRPA. This can be
achieved by taking into account coupling  with complex thermal
(e.g., two-phonon) configurations. For charge-exchange excitations
in cold nuclei this problem was resolved within the
quasiparticle-phonon nuclear model~\cite{kuz84,kuz85} and other
approaches~\cite{Drozdz}. It was found that the coupling with
complex configuration strongly affects the RPA strength
distribution. As discussed above,  the details of the GT strength
distribution are of particular importance for weak interaction rates
at low densities and temperatures. It is also desirable to extend
the approach to more microscopic effective interactions and also to
consider the case of deformed nuclei.

\begin{acknowledgments}
Discussions with Dr.~V.~A.~Kuzmin are gratefully acknowledged. This
work is supported by the Heisenberg-Landau Program, the DFG grant
(SFB 634), the Helmholtz Alliance EMMI and HIC for Fair.
\end{acknowledgments}


\begin{thebibliography}{99}
\bibitem{toro2000} M.~Di Toro, V.~Baran,
M.~Cabibbo, M.~Colonna, A.~B.~Larionov, and N.~Tsoneva, Phys. Part.
Nucl. {\bf 31}, 4333 (2000).
\bibitem{shlomo2005} S.~Shlomo and V.~M.~Kolomietz, Rep. Prog. Phys. {\bf 68},  1 (2005).
\bibitem{lang03} K.~Langanke and G.~Mart\'{i}nez-Pinedo, Rev. Mod. Phys. {\bf 75},  819 (2003).
\bibitem{coo84} J.~Cooperstein and J.~Wambach, Nucl. Phys. A {\bf 420},
591 (1984).
\bibitem{fuller} G.~M.~Fuller, W.~A.~Fowler, and M.~J.~Newman,
  Astrophys.~J. Suppl. {\bf 42},  447 (1980); {\bf 48},  279 (1982);
  Astrophys.~J. {\bf 252},  715 (1982); {\bf 293},  1 (1985).
\bibitem{auf94} M.~B.~Aufderheide, I.~Fushiki, S.~E.~Woosley, and D.~H.~Hartmann,
   Astrophys.~J. Suppl. {\bf 91},  389 (1994).
\bibitem{lan98} K.~Langanke and G.~Mart\'{i}nez-Pinedo, Phys. Lett. B {\bf 436},  19 (1998).
\bibitem{lan99} K.~Langanke and G.~Mart\'{i}nez-Pinedo, Phys. Lett. B {\bf 453},   187 (1999).
\bibitem{lan00} E.~Caurier, K.~Langanke,  G.~Mart\'{i}nez-Pinedo, and E.~Nowacki,
  Nucl. Phys. A {\bf 653},  439 (1999); K.~Langanke and G.~ Mart\'{i}nez-Pinedo,
  Nucl. Phys. A {\bf 673},  481 (2000).
\bibitem{lan01_1} K.~Langanke and G.~Mart\'{i}nez-Pinedo, At. Data Nucl. Data Tables {\bf  79},  1 (2001).
\bibitem{kar94} K.~Kar, A.~Ray, and S.~Sarkar, Astrophys. J. {\bf 434},  662 (1994).
\bibitem{nab04} J.-U.~Nabi and  H.~V.~Klapdor-Kleingrothaus,
                At. Data  Nucl. Data Tables {\bf 88},   237 (2004).
\bibitem{Paar09} N.~Paar, G.~Col\'{o}, E.~Khan, and D. Vretenar, Phys. Rev. C (to be published).
\bibitem{koo97} S.~E.~Koonin, D.~J.~Dean, K.~Langanke, Phys. Rep. {\bf 278},  1 (1997).
\bibitem{rad97} P.~B.~Radha, D.~J.~Dean, S.~E.~Koonin, K.~Langanke, and P.~Vogel,
 Phys. Rev. C {\bf 56}, 3079 (1997).
\bibitem{juodagalvis09} A.~Juodagalvis, K.~Langanke, W.~R.~Hix, G.~Martinez-Pinedo, J.~M.~Sampaio,
  \texttt{{arXiv:0909.0179v1
[nucl-th]}}.
\bibitem{hal67} J.~A.~Halbleib and R.~A.~Sorensen,  Nucl.  Phys.  A {\bf 98},   542
(1967).
\bibitem{ume75} Y.~Takahashi and H.~Umezawa, Collect. Phenom. {\bf 2},  55 (1975).
\bibitem{ume82} H.~Umezawa, H. Matsumoto and M. Tachiki,
                  {\it  Thermo field dynamics and condensed states} (North-Holland, Amstredam,
                  1982).
\bibitem{lan01_2} K.~Langanke, E.~Kolbe, and D.~J.~Dean, Phys. Rev. C {\bf 63},  032801(R)
(2001).
\bibitem{tana88} K.~Tanabe, Phys.Rev. C {\bf 37},  2802 (1988).
\bibitem{hats89} T.~Hatsuda,  Nucl. Phys. A {\bf 492},  187 (1989).
\bibitem{civi93} O. Civitarese and A.~L.~DePaoli, Z. Phys. A {\bf 344},  243 (1993).
\bibitem{kos94} D.~S. Kosov and A. I. Vdovin,  Mod. Phys. Lett. A {\bf 9},  1735 (1994).
\bibitem{KVW97} D.~S.~Kosov, A.~I.~Vdovin, and J.~Wambach, {\it in  Proceedings
of the International Conference on Nuclear Structure and Related
Topics, Dubna, 1997,} edited by S.~N.~Ershov, R.~V.~Jolos,
V.~V.~Voronov (JINR, Dubna, {E4-97-327}, 1997), p.~254.
\bibitem{oji81} I.~Ojima, Ann. Phys. {\bf 137},  1 (1981).
\bibitem{dzh08} A.~A.~Dzhioev and A.~I.~Vdovin, Int. J. Mod. Phys. E {\bf 18},  1535 (2009).
\bibitem{dzh09} A.~A.~Dzhioev, A.~I.~Vdovin, V.~Yu.~Ponomarev, and J.~Wambach,
 Yad. Phys. {\bf 72}, 1373 (2009) [Phys. At. Nucl. {\bf 72},  1320 (2009)].
\bibitem{sol92} V.~G.~Soloviev, {\it Theory of atomic nuclei: quasiparticles and phonons},
(Institute of Physics Publishing, Bristol and Philadelphia, 1992).
\bibitem{kuz84} V.~A.~Kuzmin and V.~G.~Soloviev,  J. Phys. G {\bf 10},  1507 (1984).
\bibitem{kuz85} V.~A.~Kuzmin and V.~G.~Soloviev, J. Phys. G {\bf 11},  603 (1985).
\bibitem{good81} A.~L.~Goodman,  Nucl. Phys. {\bf A352},  30 (1981).
\bibitem{civ83} O.~Civitarese, G.~G.~Dussel,  and  R.~P.~J.~Perazzo,
   Nucl. Phys. A {\bf 404},  15 (1983).
\bibitem{civ01} O.~Civitarese and M.~Reboiro, Phys. Rev. C {\bf 63},  034323 (2001).
\bibitem{heg01} A.~Heger, S.~E.~Woosley, G.~Mart\'{i}nez-Pinedo, and K.~Langanke,
    Astrophys. J. {\bf 560},  307 (2001).
\bibitem{che67} V.~A.~Chepurnov, Yad. Phys. {\bf 6}, 955 (1968) [Sov. J. Nucl. Phys. {\bf 6},  696 (1968)].
\bibitem{pom97} K.~Pomosrki, P.~Ring, G.~A.~Lalazissis, A.~Baran, Z.~Lojewski,
  B.~Nerlo-Pomorska, and M.~Warda, Nucl. Phys. A {\bf 624},  349 (1997).
\bibitem{audi93} G.~Audi and A.~H.~Wapstra, Nucl. Phys. A {\bf 565},
 1 (1993).
\bibitem{rap83} J.~Rapaport, T.~Taddeucci, T.~P.~Welch, C.~Gaarde, J.~Larsen, D.~J.~Horen,
  E.~Sugarbaker, P.~Koncz, C.~C.~Foster, C.~D.~Goodman, C.~A.~Goulding, and T.~ Masterson,
  Nucl. Phys. A {\bf 410},  371 (1983).
\bibitem{ron93} T.~T.~R\"{o}nnqvist, H.~Cond\'{e}, N.~Olsson, E.~Ramstr\"{o}m,
 R.~Zorro, J. Blomgren, A.~H{\aa}kansson, A.~Ringbom, G.~Tibell, O.~Jonsson,
  L.~Nilsson, P.-U.~Renberg, S.~Y.~van~der~Werf, W.~Unkelbach, and F.~P.~Brady,
   Nucl. Phys. A {\bf 563},  225 (1993).
\bibitem{kat94} S.~El-Kateb, K.~P.~Jackson, W.~P.~Alford, R.~Abegg, R.~E.~Azuma, B.~A.~Brown,
  A.~Celler, D.~ Frekers, O.~H\"{a}usser, R.~Helmer, R.~S.~Henderson, K.~H.~Hicks, R.~Jeppesen,
  J.~D.~King, G.~G.~Shute, B.~M.~Spicer, A.~Trudel, K.~Raywood, M.~Vetterli, and S.~Yen,
   Phys. Rev. C {\bf 49} , 3128 (1994).
\bibitem{cas76} B.~Castel, I.~Hamamoto, Phys. Lett. B {\bf 65},  27 (1976).
\bibitem{bes75} D.~R.~Bes, R.~A.~Broglia, and B.~S.~Nilsson, Phys. Rep.  {\bf 16},  1( 1975).
\bibitem{ge76occ} J.~P.~Schiffer, S.~J.~Freeman, J.~A.~Clark, C.~Deibel,
C.~R.~Fitzpatrick, S.~Gros, A.~Heinz, D.~Hirata, C.~L.~Jiang,
B.~P.~Kay, A.~Parikh, P.~D.~Parker, K.~E.~Rehm, A.~C.~C.~Villari,
V.~Werner, and C.~Wrede,  Phys. Rev. Lett. {\bf 100}, 112501 (2008).
\bibitem{Caurier} E.~Caurier, K.~Langanke, G.~Mart\'{i}nez-Pinedo, F.~Nowacki,
P.~Vogel, Phys. Lett. B {\bf 522}, 240 (2001) .
\bibitem{Lisetzki} A.~F.~Lisetskiy, E.~Caurier, K.~Langanke, G.~Mart\'{i}nez-Pinedo,
 P.~von~Neumann-Cosel, F.~Nowacki, A.~Richter, Nucl. Phys. A
{\bf 789}, 114 (2007).
\bibitem{Drozdz} S.~Dro\.zd\.z, V.~Klemt, J.~Speth, and J.~Wambach, Phys. Lett. B {\bf
166},  18 (1986); S.~Dro\.zd\.z, F.~Osterfeld, J.~Speth, J.~Wambach,
Phys. Lett. B {\bf 189},  271 (1987).

\end{thebibliography}
\end{document}